\documentclass{llncs}

\title{Stochastic Parity Games on  Lossy Channel Systems
\thanks{
Technical Report EDI-INF-RR-1416 of the School of Informatics at the University
of Edinburgh, UK. (http://www.inf.ed.ac.uk/publications/report/).
Full version (including proofs) of material presented at QEST 2013 (Buenos Aires, Argentina).
arXiv.org - CC BY 3.0.
}
}

\author{Parosh Aziz Abdulla\inst{1},
Lorenzo Clemente$^2$,
Richard Mayr$^3$,
and Sven Sandberg$^1$
}

\institute{
$^1$
Uppsala University
\quad
$^2$
LaBRI, University of Bordeaux I
\quad
$^3$
University of Edinburgh}

\usepackage{times}

\usepackage{url}

\usepackage{amsmath,amssymb,amsfonts,latexsym,stmaryrd}
\usepackage{epsfig}

\usepackage[ruled,noend,nofillcomment,linesnumbered]{algorithm2e}

\usepackage{gastex,delarray,wrapfig}
\usepackage{extarrows}
\usepackage{stmaryrd}
\usepackage{verbatim}
\usepackage[mathscr]{euscript}
\usepackage{color}
\usepackage{pifont}
\usepackage{subfigure}
\usepackage{multimedia}
\usepackage{delarray}
\usepackage{graphicx}
\usepackage{tikz}
\usepackage{listings}
\usepackage{times}
\usepackage{float}
\usepackage{pgf}
\usepackage{tikz}
\usetikzlibrary{automata}
\usetikzlibrary{trees}
\usetikzlibrary{shapes}
\usetikzlibrary{petri}
\usetikzlibrary{arrows}
\usetikzlibrary{snakes}
\usetikzlibrary{backgrounds}
\usetikzlibrary{calc}

\tikzstyle{prob-node}=[fill=blue!50!black,text=white,circle]
\tikzstyle{player-node}=[fill=blue!50!black,text=white]
\tikzstyle{transition-edge}=[->,line width=1pt]
\tikzset{background rectangle/.style={fill=yellow!10,rounded corners,draw}}

\newcommand{\hide}[1]{}

 \newcommand{\longleadsto}
    {
      \tikz \draw [->,
      line join=round,
      decorate, decoration={
        zigzag,
        segment length=4,
        amplitude=.9,post=lineto,
        post length=2pt
    }]  (0,0) -- (0.5,0);
    }

\newcommand{\lrb}[1]{[#1]}
\newcommand{\lrc}[1]{(#1)}
\newcommand{\lrd}[1]{\{#1\}}

\newcommand{\ignore}[1]{}

\newcommand{\emptyword}{\varepsilon}

\newcommand{\wpp}{w.p.p.}
\newcommand{\as}{a.s.}

\newcommand{\nat}{\mathbb N}
\newcommand{\ord}{\mathbb O}

\newcommand{\set}[1]{\lrd{#1}}
\newcommand{\setcomp}[2]{\lrd{{#1}|\;{#2}}}
\newcommand{\tuple}[1]{\lrc{#1}}
\newcommand{\denotationof}[2]{\llbracket #1\rrbracket^{#2}}
\newcommand{\game}{{\mathcal G}}
\newcommand{\tsys}{{\mathcal T}}
\newcommand{\bscc}{B}
\newcommand{\gametuple}{\tuple{\states,\zstates,\ostates,\rstates,\transition,\prob,\coloring}}
\newcommand{\selectfrom}[1]{{\it select}\lrc{#1}}
\newcommand{\restrict}{|}
\newcommand{\undef}{\bot}
\newcommand{\domof}[1]{{\it dom}\lrc{#1}}
\newcommand{\states}{S}
\newcommand{\memstates}{\states_{\memory{}}}
\newcommand{\stateset}{Q}
\newcommand{\targetset}{{\tt Target}}
\newcommand{\state}{s}

\newcommand{\zstates}{\states^0}
\newcommand{\ostates}{\states^1}
\newcommand{\rstates}{\states^R}
\newcommand{\xstates}{\states^x}
\newcommand{\ystates}{\states^{1-x}}
\newcommand{\transition}{{\longrightarrow}}

\newcommand{\tmovesto}{\longleadsto}
\newcommand{\prob}{P}

\newcommand{\parg}[1]{\paragraph{#1}\vspace{-2mm}}
\newcommand{\postof}[2]{{\it Post}_{#1}\left(#2\right)}
\newcommand{\preof}[2]{{\it Pre}_{#1}\left(#2\right)}
\newcommand{\dualpreof}[2]{\widetilde{{\it Pre}_{#1}}\left(#2\right)}

\newcommand{\complementof}[1]{\overline{#1}}
\newcommand{\gcomplementof}[2]{\stackrel{_{#1}}{_\neg}{#2}}
\newcommand{\zcut}[1]{\lrb{#1}^0}
\newcommand{\ocut}[1]{\lrb{#1}^1}
\newcommand{\zocut}[1]{\lrb{#1}^{0,1}}
\newcommand{\rcut}[1]{\lrb{#1}^R}
\newcommand{\xcut}[1]{\lrb{#1}^{x}}
\newcommand{\ycut}[1]{\lrb{#1}^{1-x}}

\newcommand{\run}{\rho}
\newcommand{\runset}{{\mathfrak R}}
\newcommand{\runsof}[1]{{\it Runs}\lrc{#1}}
\newcommand{\pth}{\pi}
\newcommand{\pthset}[1]{{\mathrm\Pi}_{#1}}
\newcommand{\strat}{f}
\newcommand{\zstrat}{\strat^0}
\newcommand{\ostrat}{\strat^1}
\newcommand{\xstrat}{\strat^x}
\newcommand{\ystrat}{\strat^{1-x}}
\newcommand{\strats}{F}
\newcommand{\zstrats}{\strats^0}
\newcommand{\ostrats}{\strats^1}
\newcommand{\xstrats}{\strats^x}
\newcommand{\ystrats}{\strats^{1-x}}
\newcommand{\xforcestrat}{{\it force}^x}
\newcommand{\yforcestrat}{{\it force}^{1-x}}

\newcommand{\yavoidstrat}{{\it avoid}^{1-x}}

\newcommand{\xallstrats}{F_{\it all}^\px}
\newcommand{\xfinitestrats}{F^\px_{\it finite}}
\newcommand{\xnomemstrats}{F^{\px}_\emptyset}
\newcommand{\yallstrats}{F_{\it all}^{\py}}
\newcommand{\yfinitestrats}{F^{\py}_{\it finite}}
\newcommand{\ynomemstrats}{F^{\py}_\emptyset}

\newcommand{\ystratset}{F^{\py}}
\newcommand{\px}{x}
\newcommand{\py}{{1-x}}
\newcommand{\pz}{0}
\newcommand{\po}{1}
\newcommand{\partialto}{\rightharpoonup}

\newcommand{\memory}{M}
\newcommand{\memconf}{m}
\newcommand{\memtrans}{\tau}
\newcommand{\memmem}{\mu}
\newcommand{\memstrattuple}{\tuple{\memory,\initmem,\memtrans,\memmem}}
\newcommand{\initmem}{\memconf_0}
\newcommand{\memstrat}[1]{\mathcal M^{#1}}
\newcommand{\memstratn}{\memstrat{}}

\newcommand{\memstratstrat}[1]{\mathit{strat}_{\memstrat{#1}}}
\newcommand{\memstratstratn}{\memstratstrat{}}

\newcommand{\om}{\omega}
\newcommand{\Om}{\Omega}
\newcommand{\probm}{{\cal P}}

\newcommand{\formula}{{\varphi}}
\newcommand{\parity}{{\it Parity}}
\newcommand{\xparity}{\mbox{$x$-}\parity}
\newcommand{\yparity}{\mbox{$(1-x)$-}\parity}

\newcommand{\always}{\Box}
\newcommand{\eventually}{\Diamond}
\newcommand{\colorset}[3]{[#1]^{\coloring#2#3}}
\newcommand{\coloring}{{\tt Col}}
\newcommand{\colorof}[1]{\coloring\lrc{#1}}

\newcommand{\winset}{W}

\newcommand{\xwinset}{\winset^x}
\newcommand{\ywinset}{\winset^{1-x}}

\newcommand{\vinset}{V}

\newcommand{\xvinset}{\vinset^x}
\newcommand{\yvinset}{\vinset^{1-x}}

\newcommand{\reachset}{{\mathcal R\,\,\!\!}}

\newcommand{\xforceset}{{\it Force}^x}
\newcommand{\yforceset}{{\it Force}^{1-x}}
\newcommand{\xavoidset}{{\it Avoid}^x}
\newcommand{\yavoidset}{{\it Avoid}^{1-x}}
\newcommand{\tp}{\alpha}
\newcommand{\btp}{\beta}

\newcommand{\cset}{{\mathcal C}}
\newcommand{\dset}{{\mathcal D}}
\newcommand{\xset}{{\mathcal X}}
\newcommand{\yset}{{\mathcal Y}}
\newcommand{\zset}{{\mathcal Z}}
\newcommand{\uset}{{\mathcal U}}
\newcommand{\vset}{{\mathcal V}}

\newcommand{\cut}{\ominus}

\newcommand{\attractor}{A}

\newcommand{\seq}[1]{\{{#1}_i\}_{i\in\ord}}

\newcommand{\statesx}{{\states^\px}}

\newcommand{\lossp}{\lambda}
\newcommand{\sglcstuple}{\tuple{\lcsstates,\lcsstatesz,\lcsstateso,\channels,\msgs,\lcstransitions,\lossp,\coloring}}
\newcommand{\sglcs}{{\mathcal L}}
\newcommand{\lcsstates}{{\tt S}}
\newcommand{\lcsstatesz}{\lcsstates^\pz}
\newcommand{\lcsstateso}{\lcsstates^\po}
\newcommand{\lcsstatesx}{\lcsstates^\px}

\newcommand{\lcsstate}{{\tt s}}
\newcommand{\msgs}{{\tt M}}
\newcommand{\msg}{{\tt m}}
\newcommand{\channels}{{\tt C}}
\newcommand{\channel}{{\tt c}}
\newcommand{\lcstransitions}{{\tt T}}
\newcommand{\lcstransition}{{\tt t}}
\newcommand{\op}{{\tt op}}
\newcommand{\nop}{{\tt nop}}

\newcommand{\chassignment}{{\tt x}}

\newcommand{\emptychannels}{\pmb{\emptyword}}

\newcommand{\transitionx}[1]{\overset{{#1}}{\transition}}

\newcommand{\ucof}[1]{{#1}\!\uparrow}
\newcommand{\ucset}{U}

\usepackage{pslatex}
\usepackage{epsfig,psfrag}
\begin{document}
\maketitle

\begin{abstract}
We give an algorithm for solving
stochastic parity games with almost-sure winning conditions on
{\it lossy channel systems}, for the case where the
players are restricted to finite-memory
strategies. 
First, we describe a general framework, where we
consider the class of 
$2\frac{1}{2}$-player games with
almost-sure parity winning conditions on possibly infinite game graphs, 
assuming that the game contains a {\it finite attractor}.
An attractor is a set of states (not necessarily absorbing)
that is almost surely re-visited regardless of the players' decisions.
We present a scheme that characterizes the set
of winning states for each player.
Then, we instantiate this scheme
to obtain an algorithm for
{\it stochastic game lossy channel systems}.
\end{abstract}

\section{Introduction}
\label{introduction:section}
\parg{Background.}
2-player games can be used to model the interaction of
a controller (player 0) who makes choices in a reactive 
system, and a malicious adversary (player 1) who represents
an attacker.
To model randomness in the system
(e.g., unreliability; randomized algorithms),
a third player `random' is defined who makes choices according
to a predefined probability distribution. The resulting 
stochastic game is called a $2\frac{1}{2}$-player game in the terminology of \cite{chatterjee03simple}.
The choices of the players induce a run of the system, and
the winning conditions of the game are expressed in terms of predicates 
on runs.

Most classic work on algorithms for stochastic games has focused
on finite-state systems (e.g.,
\cite{shapley-1953-stochastic,condon-1992-ic-complexity,AHK:FOCS98,chatterjee03simple}),
but more recently several classes of infinite-state systems have been
considered as well. 
Stochastic games on infinite-state probabilistic recursive systems (i.e.,
probabilistic pushdown automata with unbounded stacks) were studied in
\cite{Etessami:Yannakakis:ICALP05,EY:LMCS2008,EWY:ICALP08}.
A different (and incomparable) class of infinite-state systems are channel
systems, which use unbounded communication buffers instead of unbounded
recursion.

{\it Channel Systems} consist of
finite-state machines that communicate by asynchronous message passing
via unbounded FIFO communication channels. They are also known as
communicating finite-state machines (CFSM) \cite{Brand:CFSM}.

A {\it Lossy Channel System (LCS)} \cite{AbJo:lossy} consists of
finite-state machines that communicate by asynchronous message passing
via unbounded unreliable (i.e., lossy) FIFO communication channels,
i.e., messages can spontaneously disappear from channels.

A {\it Probabilistic Lossy Channel System (PLCS)}
\cite{Schnoeblen:plcs,Parosh:Alex:PLCS}
is a probabilistic variant 
of LCS where, in each computation step, messages
are lost from the channels with a given probability.
In \cite{ABDMS:FOSSACS08}, a {\em game extension} of PLCS was introduced
where the players control transitions in the control graph
and message losses are probabilistic.

The original motivation for LCS and PLCS was 
to capture the behavior of communication protocols;
such protocols are designed to operate correctly even
if the communication medium is unreliable (i.e., if messages can be lost).
However, Channel Systems (aka CFSM) are a very expressive model that
can encode the behavior of Turing machines, by storing the content of a Turing
tape in a channel \cite{Brand:CFSM}. The only reason why certain questions
are decidable for LCS/PLCS is that the message loss induces a quasi-order
on the configurations, which has the properties of a simulation.
Similarly to Turing machines and CFSM, 
one can encode many classes of infinite-state probabilistic transition
systems into a PLCS. The only requirement is that the system re-visits a 
certain finite core region (we call this an attractor; see below) with probability
one, e.g., 
\begin{itemize}
\item
Queuing systems where waiting customers in a queue drop out
with a certain probability in every time interval.
This is similar to the well-studied class of queuing systems
with impatient customers which practice {\em reneging}, i.e.,
drop out of a queue after a given maximal waiting time; see 
\cite{Wang-Li-Jiang:Review} section II.B.
Like in some works cited in \cite{Wang-Li-Jiang:Review}, the maximal waiting time
in our model is exponentially distributed.
In basic PLCS, unlike in \cite{Wang-Li-Jiang:Review}, this exponential distribution
does not depend on the current number of waiting customers.
However, an extension of PLCS with this feature would still 
be analyzable in our framework (except in the pathological case where
a high number of waiting customers increases the customers patience
exponentially, because such a system would not necessarily have a finite attractor).
\item
Probabilistic resource trading games with
probabilistically fluctuating prices.
The given stores of resources are
encoded by counters (i.e., channels), which exhibit a probabilistic decline
(due to storage costs, decay, corrosion, obsolescence).
\item
Systems modeling operation cost/reward, which is stored in counters/channels,
but probabilistically discounted/decaying over time.
\item
Systems which are periodically restarted (though not necessarily by a
deterministic schedule), due to, e.g., energy depletion or maintenance work. 
\end{itemize}
Due to this wide applicability of PLCS, we focus on this model in this paper.
However, our main results are formulated in more general terms referring
to infinite Markov chains with a finite attractor; see below.

\parg{Previous work.}

Several algorithms for symbolic model checking of PLCS have been presented 
\cite{Parosh:etal:attractor:IC,Rabinovich:plcs}.
Markov decision processes (i.e., $1\frac{1}{2}$-player games) 
on infinite
graphs induced by PLCS were studied in \cite{BBS:ACM2007},
which shows
that $1\frac{1}{2}$-player games with almost-sure
B\"uchi objectives are pure memoryless determined and decidable.
This result was later generalized to $2\frac{1}{2}$-player games
\cite{ABDMS:FOSSACS08},
and further extended 
to generalized B\"uchi objectives \cite{BS-qapl2013}.
On the other hand, $1\frac{1}{2}$-player games on 
PLCS with positive probability
B\"uchi objectives (i.e., almost-sure co-B\"uchi objectives from the
(here passive) opponent's point of view) 
can require infinite memory to win
and are also undecidable \cite{BBS:ACM2007}. 
(Undecidability and infinite memory requirement
are separate
results, since 
decidability 
does 
not imply the existence of
finite-memory strategies in infinite-state games).
If players are restricted to finite-memory strategies,
the $1\frac{1}{2}$-player game with positive probability
parity objectives
(even the more general \emph{Streett objectives})
becomes decidable \cite{BBS:ACM2007}.
Note that the finite-memory case and the infinite-memory one are a priori incomparable problems,
and neither subsumes the other.
Cf. Section~\ref{conclusions:section}.

Non-stochastic (2-player) parity games on infinite graphs were studied
in \cite{zielonka1998infinite}, where it is shown that such games are
determined, and that both players possess winning memoryless strategies in
their respective winning sets.  Furthermore,
a scheme for computing the winning
sets and winning strategies is given.
Stochastic games ($2\frac12$-player games) with parity conditions on
\emph{finite} graphs are known to be memoryless determined and
effectively solvable
\cite{alfaro-2000-lics-concurrent,chatterjee03simple,chatterjee-2006-qest-strategy}.

\parg{Our contribution.}
We give an algorithm to
decide almost-sure parity games for probabilistic lossy channel systems
in the case where the players
are restricted to finite memory strategies.
We do that in two steps.
First, we give our result in general terms
(Section~\ref{parity:section}):
We
consider the class of 
$2\frac{1}{2}$-player games with
almost-sure {\it parity} wining conditions on possibly infinite game graphs, 
under the assumption
that the game contains a {\it finite attractor}.
An attractor is a set $\attractor$ of states 
such that, regardless of the strategies used by the players,
the probability measure of the runs which
visit $\attractor$ infinitely often is one.%
\footnote{
In the game community (e.g., \cite{zielonka1998infinite})
the word {\it attractor}
is used to
denote what we  call a {\it force set}
in Section~\ref{reachability:section}.
In the infinite-state systems community
(e.g., \cite{Parosh:etal:attractor:IC,Parosh:etal:MC:infinite:journal}), the word is used 
in the same way as we use it in this paper.}
Note that this means neither that $\attractor$ is absorbing, nor
that every run must visit $\attractor$.
We present a general scheme characterizing the set
of winning states for each player.
The scheme is a non-trivial generalization of the well-known scheme for
non-stochastic games in \cite{zielonka1998infinite}
(see the remark in Section~\ref{parity:section}).
In fact, the
constructions are equivalent in the case that no probabilistic states
are present.
We show correctness of the scheme for games where each player is
restricted to a finite-memory strategy.
The correctness proof here is more involved than in the
non-stochastic case of \cite{zielonka1998infinite};
we rely on the existence of a finite attractor and the restriction
of the players to use finite-memory strategies.
Furthermore, we show that if a player is winning against all
finite-memory strategies of the other player then he
can win using a \emph{memoryless} strategy.
In the second step (Section~\ref{algorithm:section}),
we show that the scheme can be instantiated 
for  lossy channel systems.
The instantiation requires the use
of a much more involved framework than
the classical one for well quasi-ordered transition systems
\cite{Parosh:Bengt:Karlis:Tsay:general:IC}
(see the remark in Section~\ref{algorithm:section}).
The above two steps yield an algorithm to
decide parity games in the case when the players
are restricted to finite memory strategies.
If the players are allowed infinite memory, then the problem
is undecidable already for $1\frac{1}{2}$-player games
with co-B\"uchi objectives (a special case of 2-color parity objectives)
\cite{BBS:ACM2007}.
Note that even if the players are restricted to finite memory strategies,
such a strategy (even a memoryless one) on an infinite game graph
is still an infinite object. Thus, unlike for finite game graphs,
one cannot solve a game by just guessing strategies
and then checking if they are winning.
Instead, we show how to
effectively compute a finite, symbolic representation of the
(possibly infinite) set of winning states for each player
as a regular language.

\hide{
\parg{Outline.}
The next section introduces some preliminaries on stochastic games.
In Section~\ref{reachability:section}, we describe a scheme
for characterizing winning states wrt.\ reachability
objectives.
In Section~\ref{parity:section}, we describe a scheme
for parity objectives, and show its correctness under the assumptions
that the game contains a finite attractor and that
the players are restricted to finite-memory strategies.
We introduce {\it Stochastic Game Lossy Channel Systems (SG-LCS)}
together with their parity game problem in Section~\ref{sglcs:section}.
We instantiate the scheme of Section~\ref{parity:section} to
SG-LCS in Section~\ref{algorithm:section},
and then conclude in Section~\ref{conclusions:section}.
}

\section{Preliminaries}
\label{prels:section}

\parg{Notation.}
Let $\ord$ and  $\nat$ denote the set of ordinal resp.\ natural numbers.
We use $f:X\to Y$ to denote that $f$ is a total function from $X$ to $Y$, and
use $f:X\partialto Y$ to denote that $f$ is a partial function from $X$ 
to $Y$.
We write $f(x)=\bot$ to denote that $f$ is undefined on $x$, and define
$\domof{f}:=\setcomp{x}{f(x)\neq\undef}$.
We say that $f$ is an {\it extension} of $g$ if $g(x)=f(x)$ whenever $g(x)\neq\undef$.
For $X'\subseteq X$, we use $f\restrict X'$ to denote the restriction of $f$ to $X'$.
We will sometimes need to pick an arbitrary element from a set.
To simplify the exposition, we let $\selectfrom{X}$ denote an arbitrary
but fixed element of the nonempty set $X$.

A \emph{probability distribution} on a countable set $X$ is a function
$f:X\to[0,1]$ such that $\sum_{x\in X}f(x)=1$.
For a set $X$, we use  $X^*$ and $X^\omega$ to denote the sets of finite
and infinite words over $X$, respectively.
The empty word is denoted by $\emptyword$.

\parg{Games.}
A \emph{game} (of \emph{rank $n$})
is a tuple $\game=\gametuple$ defined as follows.
$\states$ is a set of \emph{states}, partitioned into
the pairwise disjoint sets of \emph{random states} $\rstates$,
states $\zstates$ of Player$~\pz$, and states $\ostates$ of Player~$\po$.
$\transition\subseteq\states\times\states$ is the
  \emph{transition relation}.
  We write $\state\transition{}\state'$ to denote that
  $\tuple{\state,\state'}\in\transition{}$.
  We assume that for each $\state$ there is at least one and at most
  countably many $\state'$ with $\state\transition{}\state'$.
The \emph{probability function}
  $\prob:\rstates\times\states\to[0,1]$ satisfies both
  $\forall\state\in\rstates.\forall\state'\in\states .
  (\prob(\state,\state')>0\iff\state\transition\state')$ and
  $\forall\state\in\rstates .\sum_{\state'\in\states}
  \prob(\state,\state') = 1$.
(The sum is well-defined since we assumed that the number of successors of any state is at most countable.)
  $\coloring:\states\to\{0,\dots,n\}$,
  where $\colorof\state$ is called the
  \emph{color} of state $s$.
Let 
  $\stateset\subseteq\states$ be a set of states.
  We use $\gcomplementof\game\stateset:=\states-\stateset$ to denote the 
  {\it complement} of $\stateset$.
  Define 
  $\zcut\stateset:=\stateset\cap\zstates$,
  $\ocut\stateset:=\stateset\cap\ostates$,
  $\zocut\stateset:=\zcut\stateset\cup\ocut\stateset$, and
  $\rcut\stateset:=\stateset\cap\rstates$.
  For  $n\in\nat$ and $\sim\;\in\{=,\leq\}$, let $\colorset
  \stateset\sim n:=\setcomp{\state\in \stateset}{\colorof\state\sim n}$
  denote the sets of states in $\stateset$ with color $\sim n$.
    A \emph{run} $\run$ in $\game$ is an infinite sequence
$\state_0\state_1\cdots$ of states s.t.
$\state_i\transition{}\state_{i+1}$ for all $i\geq 0$;
$\run(i)$ denotes $\state_i$.
A \emph{path} $\pth$ is a finite sequence $\state_0\cdots\state_n$ of
states s.t. $\state_i\transition{}\state_{i+1}$ for all $i:0\leq
i<n$.
We say that $\run$ (or $\pth$) \emph{visits} $\state$ if
$\state=\state_i$ for some $i$.
For any $\stateset\subseteq\states$, we use $\pthset{\stateset}$ to
denote the set of paths that end in some state in $\stateset$.
Intuitively, the choices of the players and the resolution of randomness induce a run
$\state_0\state_1\cdots$, 
starting in some initial state $\state_0\in\states$;
state $\state_{i+1}$ is chosen as a successor of $\state_i$,
and this choice is made by Player~0 if $\state_i\in\zstates$,
by Player~1 if $\state_i\in\ostates$,
and it is chosen randomly according to the probability distribution
$\prob(\state_i,\cdot)$ if $\state_i\in\rstates$. 
%
%

\parg{Strategies.}
For $\px\in\set{0,1}$,
a \emph{strategy} of Player~$\px$ is a partial
function $\xstrat:\pthset{\xstates}\partialto\states$ s.t.
$\state_n\transition\xstrat(\state_0\cdots\state_{n})$ if 
$\xstrat(\state_0\cdots\state_{n})$ is
defined.
The strategy $\xstrat$ prescribes for Player~$\px$ the next move,
given the current prefix of the run.
A run $\run=\state_0\state_1\cdots$ is said to be {\it consistent}
with a strategy $\xstrat$ of Player~$\px$ if 
$\state_{i+1}=\xstrat(\state_0\state_1\cdots\state_i)$
whenever $\xstrat(\state_0\state_1\cdots\state_i)\neq\bot$.
We say that $\run$ is {\it induced} by $\tuple{\state,\xstrat,\ystrat}$
if $\state_0=\state$ and $\run$ is consistent with both $\xstrat$ and $\ystrat$.
We use $\runsof{\game,\state,\xstrat,\ystrat}$ to denote the set of runs
in $\game$ induced by 
$\tuple{\state,\xstrat,\ystrat}$.
We say that $\xstrat$ is total if it is defined for every
$\pth\in\pthset{\xstates}$.
A strategy $\xstrat$ of Player~$\px$ is \emph{memoryless}
  if the next state only depends on the current state and not on the
  previous history of the run, i.e., for any path
  $\state_0\cdots\state_n\in\pthset{\xstates}$, we have
  $\xstrat(\state_0\cdots\state_n)=\xstrat(\state_n)$.

  A \emph{finite-memory strategy} updates a finite memory
  each time a transition is taken, and the next state depends only on
  the current state and memory.
  Formally, we define a \emph{memory structure} for Player~$\px$ as a quadruple
  $\memstratn=\memstrattuple$ satisfying the following properties.
  The nonempty set $\memory$ is called the \emph{memory} and
  $\memconf_0\in\memory$ is the \emph{initial memory
    configuration}.
  For a current memory configuration $m$ and a current state $s$, the next
  state is given by 
  $\memtrans : \xstates\times\memory \to\states$, where 
  $\state\transition\memtrans(\state,\memconf)$.
  The next memory configuration is given by
  $\memmem:\states\times\memory\to\memory$.
  We extend $\memmem$ to paths by
  $\memmem(\emptyword,\memconf)=\memconf$ and
  $\memmem(\state_0\cdots\state_n,\memconf) =
  \memmem(\state_n,\memmem(\state_0\cdots\state_{n-1},\memconf))$.
  The total strategy $\memstratstratn:\pthset{\xstates}\to\states$
  induced by $\memstratn$ is given by $
  \memstratstratn(\state_0\cdots\state_{n})
  :=\memtrans(\state_{n},\memmem(\state_0\cdots\state_{n-1},\initmem))
  $.
  A total strategy $\xstrat$ is said to have \emph{finite memory} if
  there is a memory structure $\memstratn=\memstrattuple$ where
  $\memory$ is finite and $\xstrat=\memstratstratn$.
Consider a run $\run=\state_0\state_1\cdots\in\runsof{\game,\state,\xstrat,\ystrat}$ where $\ystrat$ is induced by $\memstratn$.
We say that $\run$ visits the configuration
$\tuple{\state,\memconf}$ if there is an
$i$ such that $\state_i=\state$ and
$\memmem(\state_0\state_1\cdots\state_{i-1},\memconf_0)=\memconf$.
We use $\xallstrats(\game)$, $\xfinitestrats(\game)$, and $\xnomemstrats(\game)$ 
to denote the set of {\it all}, {\it finite-memory}, and
{\it memoryless} strategies respectively of Player~$\px$ in $\game$.
Note that memoryless strategies and strategies in general can be
partial, whereas for simplicity we only define total finite-memory
strategies.
 
\parg{Probability Measures.}
We use the standard definition of probability measures for a set of runs
\cite{billingsley-1986-probability}.
First, we define the measure for total strategies, and then we extend 
it to general (partial) strategies.
Let
$\Om^{\state}=\state\states^{\om}$ denote the set of all infinite
sequences of states starting from $\state$.
Consider a game $\game=\gametuple$, an initial state $\state$, and total
strategies $\xstrat$ and $\ystrat$ of Players~$\px$ and~$\py$.
For a measurable set 
${\runset}\subseteq\Om^\state$, 
we define $\probm_{\game,\state,\xstrat,\ystrat}({\runset})$ to be
the probability measure of $\runset$ under the strategies $\xstrat,\ystrat$.
This measure is well-defined
\cite{billingsley-1986-probability}.
%
%
For (partial) strategies 
$\xstrat$ and $\ystrat$ of Players~$\px$ and~$\py$,
$\sim\;\in\{<,\leq,=,\geq,>\}$,
a real number $c\in[0,1]$,
and any measurable set $\runset\subseteq\Om^\state$, 
we define $\probm_{\game,\state,\xstrat,\ystrat}({\runset}) \sim c$
iff $\probm_{\game,\state,g^\px,g^{\py}}(\runset)\sim c$ for all total
strategies $g^\px$ and $g^{\py}$ that are extensions of
$\xstrat$ resp.\ $\ystrat$.

\parg{Winning Conditions.}

The winner of the game is determined by a predicate on
infinite runs.
We assume familiarity with the syntax and semantics of the temporal
logic ${\mathit CTL}^*$ (see, e.g., \cite{CGP:book}).
Formulas are interpreted on the structure $(\states,\transition)$.
We use 
$\denotationof{\formula}{\state}$ to denote the set of runs starting from
$\state$ that satisfy the ${\mathit CTL}^*$ path-formula $\formula$.
This set is measurable \cite{Vardi:probabilistic},
and we just write
$\probm_{\game,\state,\xstrat,\ystrat}(\formula) \sim c$
instead of 
$\probm_{\game,\state,\xstrat,\ystrat}(\denotationof\formula\state) \sim c$.

We will consider games with \emph{parity}
winning conditions,
whereby Player~1 wins if the largest color that occurs infinitely often in the
infinite run is odd, and Player~0 wins if it is even.
Thus, the winning condition for Player~$\px$ can be expressed in  
${\mathit CTL}^*$ 
as 
$
 \xparity :=
 \bigvee_{i\in \{0,\dots,n\}\wedge (i\bmod 2)=x}(
 \always\eventually \colorset \states=i \wedge
 \eventually\always \colorset \states\leq i)
$.

\parg{Winning Sets.}


For a strategy $\xstrat$ of Player~$\px$,
and a set $\ystratset$ of strategies of Player~$\py$,
we define
$\xwinset(\xstrat,\ystratset)(\game,\formula^{\sim c}):=
\setcomp{\state}{\forall \ystrat\in\ystratset .
\probm_{\game,\state,\xstrat,\ystrat}(\formula)\sim c}$.
If there is a strategy $\xstrat$ such that $\state\in\xwinset(\xstrat,\ystratset)(\game,\formula^{\sim c})$,
then we say that $\state$ is a {\it winning state}
for Player~$\px$ in $\game$
wrt.\ $\formula^{\sim c}$ (and $\xstrat$ is \emph{winning at $\state$}),
provided that Player~$\py$
is restricted to strategies in
$\ystratset$.
Sometimes, when the parameters $\game$, $\state$, 
$\ystratset$,  $\formula$, and $\sim c$ are known, we will not mention them and
may simply say that ``$\state$ is a winning state'' or that
``$\xstrat$ is a winning strategy'', etc.
If 
$\state\in\xwinset(\xstrat,\ystratset)(\game,\formula^{=1})$,
then
we say that Player~$\px$
\emph{almost surely (\as)} wins from $\state$.
If
$\state\in\xwinset(\xstrat,\ystratset)(\game,\formula^{>0})$,
then
we say that Player~$\px$ wins \emph{with positive probability (\wpp)}.
We define 
$\xvinset(\xstrat,\ystratset)(\game,\formula):=
\setcomp{\state}{\forall \ystrat\in\ystratset.\;\runsof{\game,\state,\xstrat,\ystrat}\subseteq\denotationof\formula\state}$.
If $\state\in\xvinset(\xstrat,\ystratset)(\game,\formula)$, then we say that
Player~$\px$ {\it surely} wins from $\state$.
Notice that any strategy that is surely winning from a state $\state$
is also winning from $\state$ \as, i.e.,
$\xvinset(\xstrat,\ystratset)(\game,\formula)\subseteq
\xwinset(\xstrat,\ystratset)(\game,\formula^{=1})$.

\parg{Determinacy and Solvability.}
A game is called \emph{determined}, wrt. a winning condition and two sets
$\zstrats,\ostrats$ of strategies of Player~$\pz$, resp.\ Player~$\po$, if, from every state,
one of the players $\px$ has a strategy $\xstrat\in\xstrats$ that wins
against all strategies $\ystrat\in\ystrats$ of the opponent.
By \emph{solving} a determined game, we mean giving an algorithm to
compute symbolic representations of the sets of states which are 
winning for either player.
%

\parg{Attractors.}
A set $\attractor\subseteq\states$ is said
to be an {\it attractor} if,
for each state $\state\in\states$ and strategies
$\zstrat,\ostrat$ of Player~$\pz$ resp.\ Player~$\po$,
 it is the case that
$\probm_{\game,\state,\zstrat,\ostrat}(\eventually\attractor)=1$.
In other words, regardless of where
we start a run and regardless of the strategies
used by the players, we will
reach a state inside the attractor \as.
It is straightforward to see that this also implies that
$\probm_{\game,\state,\zstrat,\ostrat}(\always\eventually\attractor)=1$,
i.e., the attractor will be visited infinitely often \as.

\parg{Transition Systems.}
Consider strategies $\xstrat\in\xnomemstrats$ and 
$\ystrat\in\yfinitestrats$ of
Player~$\px$ resp.\ Player~$\py$,
where $\xstrat$ is memoryless and
$\ystrat$ is finite-memory.
Suppose that $\ystrat$ is 
induced by memory structure   $\memstratn=\memstrattuple$.
We define the {\it transition system}
$\tsys$
induced
by $\game,\ystrat,\xstrat$ to be the pair
$\tuple{\memstates,\tmovesto}$
where
$\memstates=\states\times\memory$, and
$\tmovesto\subseteq\memstates\times\memstates$ such that
$\tuple{\state_1,\memconf_1}\tmovesto\tuple{\state_2,\memconf_2}$ if
$\memconf_2=\memmem(\state_1,\memconf_1)$, and
one of the following three conditions is satisfied:
(i) 
$\state_1\in\xstates$ and
either $\state_2=\xstrat(\state_1)$ or $\xstrat(\state_1)=\bot$,
(ii)  
$\state_1\in\ystates$ and $\state_2=\memtrans(\state_1,\memconf_1)$, or
(iii) 
$\state_1\in\rstates$ and $\prob(\state_1,\state_2)>0$.
Consider the directed acyclic graph (DAG) of maximal strongly
connected components (SCCs) of the transition system 
$\tsys$.
An SCC is called a {\it bottom SCC (BSCC)} if no other SCC is 
reachable from
it.  
Observe that the existence of BSCCs is not guaranteed in an
infinite transition system.
However, if $\game$ contains a finite attractor $\attractor$ 
and $\memory$ is finite
then $\tsys$ contains at least one BSCC, and  in fact
each BSCC contains at least one element
$\tuple{\state_\attractor,\memconf}$
with $\state_\attractor\in\attractor$.
In particular, for any state $\state\in\states$,
any run $\run\in\runsof{\game,\state,\xstrat,\ystrat}$ will visit
a configuration $\tuple{\state_\attractor,\memconf}$
infinitely often \as{}
where $\state_\attractor\in\attractor$
and $\tuple{\state_\attractor,\memconf}\in\bscc$ 
for
some BSCC $\bscc$.

\section{Reachability}
\label{reachability:section}

In this section we present some concepts related
to checking reachability objectives in games.
First, we define basic notions.
Then we recall a standard scheme
(described e.g. in \cite{zielonka1998infinite})
for
checking reachability winning conditions, and
state some of its properties that we use
in the later sections.
Below, fix a  game $\game=\gametuple$.

\parg{Reachability Properties.}
  Fix a state $\state\in\states$ and sets of states 
  $\stateset,\stateset'\subseteq\states$.
Let $\postof\game{\state}:=\{\state':\state\transition\state'\}$ denote the
  set of \emph{successors} of $\state$. Extend it to sets of states  by
  $\postof\game{\stateset}:=\bigcup_{\state\in\stateset}\postof\game{\state}$.
  Note that for any given state $\state\in\rstates$,
  $\prob(\state,\cdot)$ is a probability distribution over
  $\postof\game{\state}$.
  Let $\preof\game{\state}:=\{\state':\state'\transition\state\}$ denote the
  set of \emph{predecessors} of $\state$, and extend it to sets of
  states as above.
  We define $\dualpreof\game{\stateset}:=\gcomplementof\game{\preof\game{\gcomplementof\game\stateset}}$, 
  i.e., it denotes the set of  states whose successors 
  {\it all} belong to $\stateset$.
  We say that $\stateset$ is \emph{sink-free} if
  $\postof\game{\state}\cap\stateset\neq\emptyset$ for all $\state\in\stateset$,
  and \emph{closable} if it is sink-free and 
  $\postof\game\state\subseteq\stateset$ for all $\state\in\rcut\stateset$.
  If $\stateset$ is closable then each state in $\zocut\stateset$ 
  has at least   one successor in $\stateset$, and all 
  the successors of states in $\rcut\stateset$ are  in $\stateset$.

  If $\gcomplementof\game\stateset$ is closable, we define the \emph{subgame}
    $\game\cut\stateset:=\tuple{\stateset',\zcut{\stateset'},\ocut{\stateset'},
  \rcut{\stateset'},\transition',\prob',\coloring'}$, where
  $\stateset':=\gcomplementof\game\stateset$ is the new set of states,
  $\transition':=\transition\cap({\stateset'}\times{\stateset'})$, 
  $\prob':=\prob\restrict(\rcut{\stateset'}\times{\stateset'})$, 
  $\coloring':=\coloring\restrict{\stateset'}$.
  %
  Notice that $\prob'(\state)$ is a probability 
  distribution for any $\state\in\rstates$ since ${\gcomplementof\game\stateset}$ is closable.
  We use $\game\cut\stateset_1\cut\stateset_2$ to denote 
  $(\game\cut\stateset_1)\cut\stateset_2$.

  For $\px\in\set{0,1}$, we say that $\stateset$ is an \emph{$\px$-trap} if it is closable and $\postof\game{\state}\subseteq\stateset$
  for all $\state\in\xcut\stateset$.
  Notice that $\states$ is both a $\pz$-trap and a $\po$-trap, and 
  in particular it is both sink-free and closable.
The following lemma (adapted from \cite{zielonka1998infinite}) states that, starting from a state inside a set of states 
$\stateset$  that is a trap for one player, 
the other player can surely keep the run inside $\stateset$.
\begin{lemma}
\label{trap:certainly:lemma}
If $\stateset$ is a $(\py)$-trap, then there exists
a memoryless strategy $\xstrat\in\xnomemstrats(\game)$ for Player~$\px$ 
such that
$\stateset\subseteq\xvinset(\xstrat,\yallstrats(\game))(\game,\always\stateset)$.
\end{lemma}

\parg{Scheme.}
 %
 Given a set $\targetset\subseteq\states$,  we give a scheme for 
 computing a partitioning of $\states$ into two sets
$\xforceset(\game,\targetset)$ and $\yavoidset(\game,\targetset)$
that are winning for Players~$\px$~and~$\py$.
More precisely, we define a memoryless strategy that allows
Player~$\px$ to force the game to $\targetset$ \wpp; 
and define a memoryless strategy that allows
Player~$\py$ to surely avoid $\targetset$.

First, we characterize the states that are winning for Player~$\px$,
by defining an increasing set of 
states each of which consists of winning states
for Player~$\px$, as follows:%
\ignore{
(i) 
$\reachset_0:=\targetset$;
(ii)
$\reachset_{i+1}:=\reachset_i\cup
\rcut{\preof\game{\reachset_i}}\cup
\xcut{\preof\game{\reachset_i}}\cup
\ycut{\dualpreof\game{\reachset_i}}
$ if $i+1$ is a successor ordinal; and
(iii)
$ \reachset_i:=\bigcup_{j<i}\reachset_j$ if $i>0$ is a limit ordinal.
Define
$\xforceset(\game,\targetset):=\bigcup_{i\in\ord}\reachset_i$,
and define
$\yavoidset(\game,\targetset):=\gcomplementof\game{\xforceset(\game,\targetset)}$.
}
{\small
\begin{align*}
\reachset_0&:=\targetset;\\
\reachset_{i+1}&:=\reachset_i\cup
\rcut{\preof\game{\reachset_i}}\cup
\xcut{\preof\game{\reachset_i}}\cup
\ycut{\dualpreof\game{\reachset_i}}
& & \text{ if $i+1$ is a successor ordinal; } \\
\reachset_i&:=\bigcup_{j<i}\reachset_j
& & \text { if $i>0$ is a limit ordinal; }
\end{align*}\vspace{-2mm}%
\[
\xforceset(\game,\targetset):=\bigcup_{i\in\ord}\reachset_i;\qquad
\yavoidset(\game,\targetset):=\;\gcomplementof\game{\xforceset(\game,\targetset)}.
\]}

First, we show that the iteration above converges (possibly in infinitely many steps).
To this end,
we observe that
$\reachset_i\subseteq\reachset_{i+1}$
if $i+1$ is a successor ordinal and $\reachset_j\subseteq\reachset_i$
if $j<i$ and $i$ is a limit ordinal.
Therefore $\reachset_0\subseteq\reachset_1\subseteq\cdots$.
Since the sequence is non-decreasing and since
the sequence is bounded by 
$\states$, it will eventually converge.
Define $\tp$ to be the smallest ordinal such that $\reachset_\tp=\reachset_i$ for all $i\geq \tp$.
This gives the following lemma, which also
implies that
 the $\yavoidset$ set is a trap for Player~$\px$.
(Lemmas~\ref{reachability:tp:lemma} and~\ref{not:reach:trap:lemma}
are adapted from \cite{zielonka1998infinite}, where they are stated
in a non-probabilistic setting.)
\begin{lemma}
\label{reachability:tp:lemma}
There is an $\tp\in\ord$ such that
$\reachset_\tp=\bigcup_{i\in\ord}\reachset_i$.
\end{lemma}
\begin{lemma}
\label{not:reach:trap:lemma}
$\yavoidset(\game,\targetset)$ is an $\px$-trap.
\end{lemma}
The following lemma shows correctness of the construction.
In fact, it shows that a winning player also has a
memoryless winning strategy.
\begin{lemma}
\label{reachability:correct:lemma}
There is a memoryless strategy
 $\xforcestrat(\game,\targetset)\in\xnomemstrats(\game)$ such that\\
$\xforceset(\game,\targetset)\subseteq\xwinset(\xforcestrat(\game,\targetset),\yallstrats(\game))(\game,\eventually\targetset^{>0})$; and
a memoryless strategy
$\yavoidstrat(\game,\targetset)\in\ynomemstrats(\game)$
such that\\
$\xavoidset(\game,\targetset)\subseteq
\yvinset(\yavoidstrat(\game,\targetset),\xallstrats(\game))(\game,\always(\gcomplementof\game\targetset))$.
\end{lemma}
The first claim of the lemma can be proven using
transfinite induction on $i$ to show that it  holds
for each state $\state\in\reachset_i$.
The second claim follows from 
Lemma~\ref{not:reach:trap:lemma} and
Lemma~\ref{trap:certainly:lemma}.

\section{Parity Conditions}
\label{parity:section}
We describe a scheme for solving stochastic parity games with almost-sure 
winning conditions on infinite graphs,
under the conditions that the game has a finite attractor (as
defined in Section~\ref{prels:section}),
and that the players are restricted to finite-memory strategies.

By induction on $n$, we define two sequences of functions
$\cset_0,\cset_1,\ldots$ and $\dset_0,\dset_1,\ldots$ s.t.,
for each $n \geq 0$ and game $\game$ of rank at most $n$,
$\cset_n(\game)$ characterizes the states from which Player~$\px$ is winning \as, where $\px=n \bmod 2$,
and $\dset_n(\game)$ characterizes the set of states from which Player~$\px$ is winning \wpp{}.
The scheme for $\cset_n$ is related to \cite{zielonka1998infinite};
cf. the remark at the end of this section.
In both cases,
we provide a memoryless strategy that is winning for Player~$\px$;
Player~$\py$ is always restricted to finite-memory.

For the base case, let $\cset_0(\game):=\states$ and $\dset_0(\game):=\states$
for any game $\game$ of rank 0.
Indeed, from any configuration Player 0 trivially wins \as/\wpp{} because there is only color 0.

For $n\geq 1$, let $\game$ be a game of rank $n$.
$\cset_n(\game)$ is defined with the help of two auxiliary transfinite sequences
$\set{\xset_i}_{i\in\ord}$ and $\set{\yset_i}_{i\in\ord}$.
The construction ensures that
$\xset_0\subseteq\yset_0\subseteq\xset_1\subseteq\yset_1\subseteq\cdots$, 
and that the elements of $\xset_i,\yset_i$ are winning \wpp{}
for Player~$\py$.
The construction alternates as follows.
%
%
In the inductive step, we have already constructed
$\xset_j$ and $\yset_j$ for all
$j<i$.
Our construction of $\xset_j$ and $\yset_j$ is in three steps:
\begin{enumerate}
\item $\xset_i$ is the set of states
  where Player~$\py$ can force the run to visit
  $\bigcup_{j<i}\yset_j$
  \wpp.
\item Find a set of states where
  Player~$\py$ wins \wpp{} in
  $\game\cut\xset_i$.
\item Take $\yset_i$ to be the union of $\xset_j$ and the set constructed in step 2.
\end{enumerate}
We next show how to find the winning states in 
$\game\cut\xset_i$ in step 2.
%
\ignore{
We first compute the set of states where Player~$\py$ can force the run
to reach $\yset_j$ \wpp{} for some $j<i$.
We call this set $\xset_i$.
The set $\xset_i$ is winning \wpp{} for 
 Player~$\py$, given
the induction hypothesis that all $\yset_j$ are winning.
Then, we find a set of states where  Player~$\py$ wins \wpp{} in
$\game\cut\xset_i$.
The union of this set and $\xset_i$ is called $\yset_i$; and we will see that
$\yset_i$ is also winning \wpp{} for Player~$\py$ in $\game$.
The next task is to find the winning states in 
$\game\cut\xset_i$.
}%
%
We first compute the set of states where Player~$\px$ can force the play
in $\game\cut\xset_i$ to reach a state with color $n$ 
\wpp.
We call this set $\zset_i$.
The subgame $\game\cut\xset_i\cut\zset_i$ does not contain any states of color $n$.
Therefore, this game can be completely solved, using the already constructed
function $\dset_{n-1}(\game\cut\xset_i\cut\zset_i)$.
We will prove that the states where Player~$\py$ wins \wpp{} in 
$\game\cut\xset_i\cut\zset_i$ are winning
\wpp{} also in $\game$.
We thus take $\yset_i$ as the union of $\xset_i$ and 
$\dset_{n-1}(\game\cut\xset_i\cut\zset_i)$.
We define the sequences formally:
\ignore{
$\xset_i:=\yforceset(\game,\bigcup_{j<i}\yset_j)$,
$\zset_i:=\xforceset(\game\cut\xset_i,\colorset{\gcomplementof\game\xset_i}=n)$,
$\yset_i:=\xset_i\cup\dset_{n-1}(\game\cut\xset_i\cut\zset_i)$,
and
$\cset_n(\game):=\gcomplementof\game{(\bigcup_{i\in\ord}\xset_i)}$.
}
\begin{align*}
  \begin{array}{rl}
    \xset_i&:=\yforceset(\game,{\textstyle\bigcup_{j<i}\yset_j}),\\
    \zset_i&:=\xforceset(\game\cut\xset_i,\colorset{\gcomplementof\game\xset_i}=n),\\
    \yset_i&:=\xset_i\cup\dset_{n-1}(\game\cut\xset_i\cut\zset_i),
  \end{array}
  \qquad \qquad
  \cset_n(\game):=\gcomplementof\game{\textstyle(\bigcup_{i\in\ord}\xset_i)}.
\end{align*}
%
Notice that the subgames $\game\cut\xset_i$ and
$\game\cut\xset_i\cut\zset_i$ are well-defined since
(by Lemma~\ref{not:reach:trap:lemma}) 
$\gcomplementof\game\xset_i$ is closable in $\game$, and 
$\gcomplementof{\game\cut\xset_i}\zset_i$ is closable in $\game\cut\xset_i$.

We now construct $\dset_n(\game)$.
Assume that we can construct $\cset_n(\game)$.
We will define the transfinite sequence $\seq \uset$ and the auxiliary
transfinite sequence $\seq \vset$.
We again precede the formal definition with an informal explanation of
the idea.
The construction ensures that $\uset_0\subseteq \vset_0\subseteq \uset_1\subseteq
\vset_1\subseteq\cdots$, and that all $\uset_i$, $\vset_i$ are winning
\wpp{} for Player~$\px$ in $\game$.
The construction alternates in a similar manner to the construction of
$\cset_n$.
In the inductive step, we have already constructed $\vset_j$ for all $j<i$.
We first compute the set of states where Player~$\px$ can force the
play to reach $\vset_j$ \wpp{} for some $j<i$.
We call this set $\uset_i$.
It is clear that $\uset_i$ is winning \wpp{}
for Player~$\px$ in $\game$, given
the induction hypothesis that all $\vset_j$ are winning.
Then, we find a set of states where Player~$\px$ wins 
\wpp{} in
$\game\cut\uset_i$.
It is clear that $\cset_n(\game\cut\uset_i)$ is such a set.
This set is winning \wpp{} for Player~$\px$, 
because a play starting in
$\cset_n(\game\cut\uset_i)$ either stays in this set and 
Player~$\px$ wins with
probability 1, or the play leaves $\cset_n(\game\cut\uset_i)$ 
and enters $\uset_i$ which,
as we already know, is winning \wpp.
We thus take $\vset_i$ as the union of $\uset_i$ and $\cset_n(\game\cut\uset_i)$.
We define the sequences formally by
\ignore{
$\uset_i:=\xforceset(\game,\bigcup_{j<i}\vset_j)$,
$\vset_i:= \uset_i\cup\cset_n(\game\cut\uset_i)$, and
$\dset_n(\game):=\bigcup_{i\in\ord}\uset_i$.}%
\[
  \begin{array}{rl}
    \uset_i&:=\xforceset(\game,\textstyle{\bigcup_{j<i}\vset_j}),\\
    \vset_i&:= \uset_i\cup\cset_n(\game\cut\uset_i),
  \end{array}
  \qquad \qquad
  \dset_n(\game):=\textstyle{\bigcup_{i\in\ord}\uset_i}.
\]
%
By the definitions, for $j < i$ we get
$\yset_j \subseteq \xset_i \subseteq \yset_i$ and
$\vset_j \subseteq \uset_i \subseteq \vset_i$.
As in Lemma~\ref{reachability:tp:lemma},
we can prove that these sequences converge.
\begin{lemma}
\label{tp:lemma}
There are $\tp,\btp\in\ord$ such that
(i) $\xset_\tp = \yset_\tp = \bigcup_{i\in\ord}\yset_i$,
(ii) $\cset_n(\game)=\gcomplementof\game\xset_\tp$,
(iii) $\uset_\btp = \vset_\btp = \bigcup_{i\in\ord}\vset_i$, and
(iv) $\dset_n(\game)=\uset_\btp$.
\end{lemma}
The following lemma shows the correctness of
the construction.
Recall that we assume that $\game$ is of rank $n$ and that it contains
a finite attractor.
Let $x=n \bmod 2$.
\begin{lemma}
\label{cn:complemenmt:lemma}
There are memoryless strategies
$\xstrat_c,\xstrat_d,\in\xnomemstrats(\game)$ and
$\ystrat_c,\ystrat_d\in\ynomemstrats(\game)$
such that the following properties hold:\\
(i)
$\cset_n(\game)\;\subseteq\;\xwinset(\xstrat_c,\yfinitestrats(\game))(\game,{\xparity}^{=1} )$.\\
(ii)
$\gcomplementof\game{\cset_n(\game)}\;\subseteq\;\ywinset(\ystrat_c,\xfinitestrats(\game))(\game,{\yparity}^{>0} )$.\\
(iii)
$\dset_n(\game)\;\subseteq\;\xwinset(\xstrat_d,\yfinitestrats(\game))(\game,{\xparity}^{>0} )$.\\
(iv)
$\gcomplementof\game{\dset_n(\game)}\;\subseteq\;\ywinset(\ystrat_d,\xfinitestrats(\game))(\game,{\yparity}^{=1} )$.
\end{lemma}
\begin{proof}
Using induction on $n$,
we define the strategies $\xstrat_c,\xstrat_d,\ystrat_c,\ystrat_d$, 
and prove that the strategies are indeed winning.

\parg{$\xstrat_c$.}
For $n\geq 1$, let $\tp$ be as defined in Lemma~\ref{tp:lemma}.
Let $\complementof{\xset_\tp}:=\gcomplementof\game{\xset_\tp}$ and 
$\complementof{\zset_\tp}:=\gcomplementof\game{\zset_\tp}$.
We know that $\cset_n(\game)=\complementof{\xset_\tp}$.
For a state $\state\in\cset_n(\game)$, we define $\xstrat_c(\state)$ depending
on the membership of $\state$ in one of the following three partitions
of $\cset_n(\game)$:
(1) $\complementof{\xset_\tp}\cap\complementof{\zset_\tp}$,
(2) $\complementof{\xset_\tp}\cap\colorset{\zset_\tp}< n$, and
(3) $\complementof{\xset_\tp}\cap\colorset{\zset_\tp}=n$.

\begin{enumerate}
\item
$\state\in \complementof{\xset_\tp}\cap\complementof{\zset_\tp}$.
Define $\game':=\game\cut\xset_\tp\cut\zset_\tp$.
From Lemma~\ref{tp:lemma}, we have that
$\xset_{\tp+1}-\xset_\tp=\emptyset$.
By the construction of $\yset_i$ we have, for arbitrary $i$,
that $\dset^{n-1}(\game\cut\xset_i\cut\zset_i)=\yset_i-\xset_i$,
and by the construction of $\xset_{i+1}$, we have that
$\yset_i-\xset_i\subseteq\xset_{i+1}-\xset_i$.
By combining these facts we obtain
$\dset^{n-1}(\game')\subseteq\xset_{\tp+1}-\xset_\tp=\emptyset$.
Since $\game\cut\xset_i\cut\zset_i$ does not contain any states of
color $n$
(or higher), it follows by the induction hypothesis that there is a
memoryless strategy $\strat_1\in\xnomemstrats(\game')$ such that
$\gcomplementof{\game'}{\dset_{n-1}(\game')}\;\subseteq\;\xwinset(\strat_1,\yfinitestrats(\game'))(\game',{\xparity}^{=1} )$.
We define $\xstrat_c(\state):=\strat_1(\state)$.
\item
$\state\in \complementof{\xset_\tp}\cap\colorset{\zset_\tp}< n$.
Define $\xstrat_c(\state):=\xforcestrat(\game\cut\xset_\tp,\colorset{\zset_\tp}=n)(\state)$.

\item
$\state\in \complementof{\xset_\tp}\cap\colorset{\zset_\tp}=n$.
By  Lemma~\ref{not:reach:trap:lemma} we know that 
$\postof\game\state\cap\complementof{\xset_\tp}\neq\emptyset$.
Define 
$\xstrat_c(\state):=\selectfrom{\postof\game\state\cap\complementof{\xset_\tp}}$.
\end{enumerate}
Let $\ystrat\in\yfinitestrats(\game)$ be a finite-memory strategy for Player~$\py$.
We show that\\
$\probm_{\game,\state,\xstrat_c,\ystrat}(\xparity )=1$
for any state $\state\in\cset_n(\game)$.
First, we show that, any run 
$\state_0\state_1\cdots\in\runsof{\game,\state,\xstrat_c,\ystrat}$ will always stay inside
$\complementof{\xset_\tp}$, i.e.,
 $\state_i\in\complementof{\xset_\tp}$ for all $i\geq 0$.
We use induction on $i$.
The base case follows from $\state_0=\state\in\complementof{\xset_\tp}$.
For the induction step, we assume that
$\state_i\in\complementof{\xset_\tp}$, and show that
$\state_{i+1}\in\complementof{\xset_\tp}$.
We consider the following cases:
\begin{itemize}
\item
$\state_i\in\ycut{\complementof{\xset_\tp}}\cup\rcut{\complementof{\xset_\tp}}$.
The result follows
since $\complementof{\xset_\tp}$ is a
($\py$)-trap in $\game$
(by Lemma~\ref{not:reach:trap:lemma}).
\item 
$\state_i\in\xcut{\complementof{\xset_\tp}\cap\complementof{\zset_\tp}}$.
We know that $\state_{i+1}=\strat_1(\state_i)$.
Since $\strat_1\in\xnomemstrats(\game\cut\xset_\tp\cut\zset_\tp)$ it follows that
$\state_{i+1}\in\complementof{\xset_\tp}\cap\complementof{\zset_\tp}$ and in particular
$\state_{i+1}\in\complementof{\xset_\tp}$.
\item
$\state_i\in\xcut{\complementof{\xset_\tp}\cap\colorset{\zset_\tp}< n}$.
We know that $\state_{i+1}=\xforcestrat(\game\cut\xset_{\tp},\colorset{\zset_\tp}=n) (\state_i)$.
The result follows by the fact that
$\xforcestrat(\game\cut\xset_{\tp},\colorset{\zset_\tp}=n)$ is a strategy
 in $\game\cut\xset_\tp$.
\item
$\state_i\in\xcut{\complementof{\xset_\tp}\cap\colorset{\zset_\tp}=n}$.
We have $\state_{i+1}\in
\postof\game{\state_i}\cap\complementof{\xset_\tp}$
and in particular $\state_{i+1}\in\complementof{\xset_\tp}$.
\end{itemize}
Let us again consider a run 
$\run\in\runsof{\game,\state,\xstrat,\ystrat}$.
We show that $\run$ is \as{} winning for Player~$\px$
with respect to $\xparity$ in $\game$.
Let $\ystrat$ be induced by a memory structure
$\memstratn=\memstrattuple$.
Let $\tsys$ be the transition system induced
by $\game$, $\xstrat$, and $\ystrat$.
As explained in Section~\ref{prels:section},
$\run$ will \as{} visit
a configuration $\tuple{\state_\attractor,\memconf}\in\bscc$ 
for some BSCC $\bscc$ in $\tsys$.
This implies that 
each state that occurs in $\bscc$ will \as{}
be visited infinitely often by $\run$.
There are two possible cases:
(i) There is a configuration $\tuple{\state_\bscc,\memconf}\in\bscc$ 
with $\coloring(\state_\bscc)=n$.
Since each state in $\game$ has color at most $n$, 
Player~$\px$ will \as{} win.
(ii)  There is no configuration 
$\tuple{\state_\bscc,\memconf}\in\bscc$ 
with $\coloring(\state_\bscc)=n$.
This implies that
$\setcomp{\state_\bscc}{\tuple{\state_\bscc,\memconf}\in\bscc}
\subseteq\complementof\zset$, and hence
Player~$\px$ uses the strategy
$\strat_1$ to win the game.
%
%

\parg{$\ystrat_c$.}


We define a strategy $\ystrat_c$ such that
$\xset_i\subseteq\yset_i\subseteq\ywinset(\ystrat_c,\xfinitestrats(\game))(\game,{\yparity}^{>0})$ for all $i$.
The result follows then from the definition of $\cset_n(\game)$.
The inclusion $\xset_i\subseteq\yset_i$ holds by the definition of $\yset_i$.
For any state $\state\in\overline{\cset_n(\game)}$,
we define $\ystrat_c(\state)$ as follows.
Let $\beta$ be the smallest ordinal such that $s\in\yset_\beta$.
Such a $\beta$ exists by the well-ordering of ordinals
and since $\overline{\cset_n(\game)}=\bigcup_{i\in\ord}\xset_i=\bigcup_{i\in\ord}\yset_i$.
Now there are two cases:
\begin{itemize}
\item
  $\state\in\xset_\beta-\bigcup_{j<\beta}\yset_j$.
  Define
  $\ystrat_c(\state):=
  f_1(\state):=
  \yforcestrat(\game,\bigcup_{j<\beta}\yset_j)(\state)$.

\item
  $\state\in\dset_{n-1}(\game\cut\xset_\beta\cut\zset_\beta)$.
  By the induction hypothesis (on $n$),
  there is a memoryless strategy
  $\strat_2\in\ynomemstrats(\game)$ of Player~$\py$ such that
  $\state\in
  \ywinset(\strat_2,\xfinitestrats(\game\cut\xset_\beta\cut\zset_\beta))
  (\game\cut\xset_\beta\cut\zset_\beta,{\yparity}^{>0} )$.
  Define $\ystrat_c(\state):=\strat_2(\state)$.
\end{itemize}
Let $\xstrat\in\xfinitestrats(\game)$ be an arbitrary finite-memory strategy for
Player~$\px$.
We now use induction on $i$ to show that
$\probm_{\game,\state,\ystrat_c,\xstrat}(\yparity )>0$
for any state $\state\in\yset_i$.
There are three cases:

\begin{enumerate}
\item
  If $\state\in\bigcup_{j<i}\yset_j$ then the result follows
  by the induction hypothesis (on $i$).
\item
  If $\state\in\xset_i - \bigcup_{j<i}\yset_j$ then we know that
  Player~$\py$, can use $\strat_1$ to force the game to
  $\bigcup_{j<i}\yset_j$
  from which she wins \wpp.
\item
  If $\state\in\dset_{n-1}(\game\cut\xset_i\cut\zset_i)$ then
  Player~$\py$ uses $\strat_2$.
  There are now two sub-cases:
  either (i) there is a run from $s$
  consistent with $\xstrat$ and $\ystrat_c$
  that reaches $\xset_i$;
  or (ii) there is no such run.
  In sub-case (i), the run reaches $\xset_i$ \wpp{}
  and then by cases~1 and~2 Player~$\py$ wins \wpp.
  In sub-case (ii), any run stays forever outside $\xset_i$.
  So the game is in effect played on $\game\cut\xset_i$.
  Notice then that any run from $\state$ that is consistent
  with $\xstrat$ and $\ystrat_c$
  stays forever in $\game\cut\xset_i\cut\zset_i$.
  The reason is that (by Lemma~\ref{not:reach:trap:lemma})
  $\gcomplementof{\game\cut\xset_i}\zset_i$
  is an $\px$-trap in
  $\game\cut\xset_i$.
  Since any run remains inside $\game\cut\xset_i\cut\zset_i$,
  Player~$\py$ wins \wpp{} wrt.\ $\yparity$ using $\strat_2$.
\end{enumerate}

\parg{$\xstrat_d$.}

For any state $\state$,
let $\beta$ be the smallest ordinal such that $s\in\yset_\beta$.
We define $\xstrat_d(\state)$ by two cases:
\begin{itemize}
\item
  $\state\in\uset_\beta-\bigcup_{j<\beta}\vset_j$.
  Define
  $\xstrat_d(\state):=\strat_1(\state):=\xforcestrat(\game,\bigcup_{j<\beta}\vset_j)(\state)$.
\item
  $\state\in\cset_n(\game\cut\uset_\beta)$.
  By the induction hypothesis (on $n$),
  Player~$\px$ has a winning memoryless strategy $\strat_2$ inside
  $\game\cut\uset_i$.
  Define $\xstrat_d(\state):=\strat_2(\state)$.
\end{itemize}
Let $\ystrat\in\yallstrats(\game)$ be an arbitrary strategy for
Player~$\py$.
We now use induction on $i$ to show that
$\probm_{\game,\state,\xstrat_d,\ystrat}(\xparity )>0$
for any state $\state\in\vset_i$.
There are three cases:
\begin{enumerate}
\item
  If $\state\in\bigcup_{j<i}\vset_j$ then the result follows
  by the induction hypothesis (on $i$).
\item
  If $\state\in\uset_i - \bigcup_{j<i}\vset_j$ then we know that
  Player~$\px$ can use $\strat_1$ to force the game to
  $\bigcup_{j<i}\vset_j$
  from which she wins \wpp{} by the previous case.
\item
  If $\state\in\cset_n(\game\cut\uset_i)$ then
  Player~$\px$ uses $\strat_2$.
  There are now two sub-cases:
  either (i) there is a run from $s$
  consistent with $\xstrat_d$ and $\ystrat$
  that reaches $\uset_i$;
  or (ii) there is no such run.
  In sub-case (i), the run reaches $\uset_i$ \wpp{}
  and then by cases~1 and~2 Player~$\px$ wins \wpp.
  In sub-case (ii), any run stays forever outside $\uset_i$.
  Hence,
  Player~$\px$ wins \as{} wrt.\ $\xparity$ using $\strat_2$.
\end{enumerate}

\ignore{
We define $\xstrat_d$ by induction on $i$.
There are two possible cases.
If $\state\in\uset_i$ then define
$\xstrat_d(\state):=\xforcestrat(\game,\bigcup_{j<i}\vset_j)(\state)$.
This strategy allows Player~$\px$ to reach $\bigcup_{j<i}\vset_j$ 
\wpp{} from
which it wins \wpp{} by the
induction hypothesis (induction on $i$).
If $\state\in\cset_n(\game\cut\uset_i)$, then 
(by induction on $n$)
Player~$\px$ has a winning memoryless strategy $\strat$ inside
$\game\cut\uset_i$.
If a run $\run$ starting from $\state\in\cset_n(\game\cut\uset_i)$  
stays inside $\game\cut\uset_i$, then
 Player~$\px$ uses $\strat$ to win the game.
Otherwise, $\run$ enters $\uset_i$ from which it wins as described above.
}%

\parg{$\ystrat_d$.} %
By the definition of $\uset_i$ we know that
$\bigcup_{j<i}\vset_j\subseteq\uset_i$, and by the definition 
of $\vset_i$ we know that $\uset_i\subseteq\vset_i$.
Thus,
$\uset_0\subseteq\vset_0\subseteq
\uset_1\subseteq\vset_1\subseteq\cdots$, and hence there is an $\tp\in\ord$
such that $\uset_i=\vset_i=\uset_\tp$ for all $i\geq\tp$.
This means that $\dset_n(\game)=\uset_\tp$ and hence by
 Lemma~\ref{not:reach:trap:lemma}
we know that $\gcomplementof\game\dset_n(\game)$ is an $\px$-trap.
Furthermore, since $\vset_\tp=\uset_\tp\cup\cset_n(\game\cut\uset_\tp)$,
where the union is disjoint,
it follows that $\cset_n(\game\cut\uset_\tp)=\emptyset$ and hence,
by the induction hypothesis,
Player~$\py$ has a memoryless strategy $\strat\in\ynomemstrats(\game)$
that is winning \wpp{} against all finite memory
strategies $\xstrat\in\xfinitestrats(\game)$
on all states in
$\gcomplementof\game\uset_\tp=\gcomplementof\game\dset_n(\game)$.
Below, we show that $\strat$ indeed allows Player~$\py$ to win
almost surely.

Fix a finite-memory strategy $\xstrat\in\xfinitestrats(\game)$.
Let $\xstrat$ be induced by a memory structure
$\memstratn=\memstrattuple$.
Consider a run
$\run\in\runsof{\game,\state,\strat,\xstrat}$.
Then, $\run$ will surely stay inside $\game\cut\uset_\tp$.
The reason is that
$\gcomplementof\game\uset_\tp$ is a trap for Player~$\px$
by  Lemma~\ref{not:reach:trap:lemma}, and that
$\strat$ is a strategy defined inside
$\game\cut\uset_\tp$.
Let $\tsys$ be the transition system induced
by $\game$, $\xstrat$, and $\strat$.
As explained in Section~\ref{prels:section},
$\run$ will \as{} visit
a configuration $\tuple{\state_\attractor,\memconf}\in\bscc$ 
for some BSCC $\bscc$ in $\tsys$.
This implies that 
each configuration in $\bscc$ will \as{}
be visited infinitely often by $\run$.
Let $n$ be the maximal color occurring among
the states of $\bscc$.
Then, either 
(i) $n\bmod 2=\px$ in which case
all states inside $\bscc$ are almost sure losing for
Player~$\py$; or
(ii) $n\bmod 2=1-\px$ in which case
all states inside $\bscc$ are almost sure winning for Player~$\py$.
The result follows from the fact that case (i) gives a contradiction
since all states in 
$\gcomplementof\game\uset_\tp=\gcomplementof\game\dset_n(\game)$
(including those in $\bscc$) are winning for Player~$\py$ \wpp.
Define $\ystrat_d(\state):=\strat(\state)$.

\end{proof}
The following theorem  follows immediately from the  previous lemmas.
\begin{theorem}
  Stochastic parity games with almost sure winning conditions on
  infinite graphs are memoryless determined, provided there exists a
  finite attractor and the players are restricted to finite-memory
  strategies.

\ignore{
  In other words, the set of states can be partitioned into three
  sets: states where player 0 wins \as, where both players can
  guarantee winning \wpp{}, and where player 1 wins \as.
  Player 0 possesses a memoryless strategy that ensures winning \as{}
  in the first partition and winning \wpp{} in the second partition,
  against arbitrary finite-memory strategies of player 1.
  Player 1 possesses a memoryless strategy that ensures winning \wpp{}
  in the second partition and winning \as{} in the third partition,
  against arbitrary finite-memory strategies of player 0.
}
\end{theorem}

\parg{Remark.}
The scheme for $\cset_n$ is adapted from the well-known scheme for
non-stochastic games in \cite{zielonka1998infinite}; in fact, the
constructions are equivalent in the case that no probabilistic states
are present.
Our contribution \emph{to the scheme} is: (1) $\cset_n$ is a
non-trivial extension of the scheme in \cite{zielonka1998infinite} to
handle probabilistic states; (2) we introduce the alternation between
$\cset_n$ and $\dset_n$; (3) the construction of $\dset_n$ is new and
has no counterpart in the non-stochastic case of
\cite{zielonka1998infinite}.

\section{Lossy Channel Systems}
\label{sglcs:section}
 
A \emph{lossy channel system (LCS)} \cite{AbJo:lossy}
is a finite-state machine
equipped with a finite number of unbounded fifo channels (queues).
The system is \emph{lossy} in the sense that, before and after a transition, an
arbitrary number of messages may be lost from the channels.
We consider \emph{stochastic game-LCS (SG-LCS)}: each individual %
message is lost independently with probability $\lossp$ 
in every step, where
$\lossp >0$ is a parameter of the system.
%
The set of states is 
partitioned into states belonging to Player~$\pz$ and~$\po$.
The player who owns the current control-state chooses an enabled
outgoing transition. 
Formally, a SG-LCS of rank $n$ is a tuple 
 $\sglcs=\sglcstuple$ where $\lcsstates$ is a
finite set of \emph{control-states} partitioned into states
$\lcsstatesz,\lcsstateso$ of Player~$\pz$ 
and~$\po$; 
$\channels$ is a finite set of \emph{channels}, $\msgs$ is a finite
set called the \emph{message alphabet}, $\lcstransitions$ is a set of
\emph{transitions}, $0<\lossp<1$ is the \emph{loss rate}, and
$\coloring:\states\to\{0,\dots,n\}$ is the {\it coloring} function.
Each transition $\lcstransition\in\lcstransitions$ is of the form
$\lcsstate\transitionx{\op}\lcsstate'$, where
$\lcsstate,\lcsstate'\in\lcsstates$ and $\op$ is one of
the following three forms:
$\channel!\msg$ (send message $\msg\in\msgs$ in channel
$\channel\in\channels$), $\channel?\msg$ (receive message $\msg$ from
channel $\channel$), or $\nop$ (do not modify the channels).
The SG-LCS $\sglcs$ induces a game 
$\game=\gametuple$, where
$\states=\lcsstates\times(\msgs^*)^\channels\times\{0,1\}$.
That is, each state in the game consists of a control-state, a function
that assigns a finite word over the message alphabet to each channel,
and one of the symbols 0 or 1.
States where the last symbol is 0 are random:
$\rstates=\lcsstates\times(\msgs^*)^\channels\times\{0\}$.
The other states belong to a player according to the control-state:
$\statesx=\lcsstatesx\times(\msgs^*)^\channels\times\{1\}$.
Transitions out of states of the form
$\state=(\lcsstate,\chassignment,1)$ model transitions in 
$\lcstransitions$ leaving state $\lcsstate$.
On the other hand, transitions leaving states of the form
$\state=(\lcsstate,\chassignment,0)$ model message losses.
If $\state=(\lcsstate,\chassignment,1),
\state'=(\lcsstate',\chassignment',0)\in\states$, then there is a
transition $\state\transition\state'$ in the game iff one of the
following holds:
(i)
$\lcsstate\transitionx{\nop}\lcsstate'$ and
  $\chassignment=\chassignment'$;
(ii)
$\lcsstate\transitionx{\channel!\msg}\lcsstate'$,
  $\chassignment'(\channel)=\chassignment(\channel)\msg$, and for all
  $\channel'\in\channels-\{\channel\}$,
  $\chassignment'(\channel')=\chassignment(\channel')$; and
(iii)
 $\lcsstate\transitionx{\channel?\msg}\lcsstate'$,
  $\chassignment(\channel)=\msg\chassignment'(\channel)$, and for all
  $\channel'\in\channels-\{\channel\}$,
  $\chassignment'(\channel')=\chassignment(\channel')$.
Every state of the form $(\lcsstate,\chassignment,0)$ has at least one
successor, namely $(\lcsstate,\chassignment,1)$.
If a state $(\lcsstate,\chassignment,1)$ does not have successors
according to the rules above, then we add a transition
$(\lcsstate,\chassignment,1)\transition(\lcsstate,\chassignment,0)$,
to ensure that the induced game is sink-free.
%
To model message losses, we introduce the subword ordering $\preceq$
on words: $x\preceq y$ iff $x$ is a word obtained by removing zero or
more messages from arbitrary positions of $y$.
This is extended to channel states
$\chassignment,\chassignment':\channels\to\msgs^*$ by
$\chassignment\preceq\chassignment'$ iff
$\chassignment(\channel)\preceq\chassignment'(\channel)$ for all
channels $\channel\in\channels$, and to game states
$\state=(\lcsstate,\chassignment,i),\state'=(\lcsstate',\chassignment',i')\in\states$
by $\state\preceq\state'$ iff $\lcsstate=\lcsstate'$,
$\chassignment\preceq\chassignment'$, and $i=i'$.
For any $\state=(\lcsstate,\chassignment,0)$ and any $\chassignment'$
such that $\chassignment'\preceq\chassignment$, there is a transition
$\state\transition(\lcsstate,\chassignment',1)$.
The probability of random transitions is given by
$\prob((\lcsstate,\chassignment,0),(\lcsstate,\chassignment',1)) =
a\cdot\lossp^b\cdot(1-\lossp)^c$, where $a$ is the number of ways to
obtain $\chassignment'$ by losing messages in $\chassignment$, $b$ is
the total number of messages needed to be lost in all channels
in order to obtain $\chassignment'$ from $\chassignment$, and $c$ is the
total number of messages in all channels of $\chassignment'$
(see \cite{Parosh:etal:attractor:IC} for details).
Finally, for a state $\state=(\lcsstate,\chassignment,i)$, we define
$\coloring(\state):=\coloring(\lcsstate)$.
Notice that the graph of the game is bipartite, in the sense that
a state  in $\rstates$ has only
transitions to states in $\zocut\states$,
and vice versa.

In the qualitative {\it parity game problem} for SG-LCS, we 
want to characterize the sets of configurations
where Player~$\px$ can force the \xparity{} condition to hold \as,
for both players.

\section{From Scheme to Algorithm}
\label{algorithm:section}
We transform the scheme of Section~\ref{parity:section}
into an algorithm for deciding the \as{} parity game problem for SG-LCS.
Consider an SG-LCS $\sglcs=\sglcstuple$ and
the induced game $\game=\gametuple$
of some rank $n$.
Furthermore,  assume that the players
are restricted to finite-memory strategies.
We show the following.
\begin{theorem}
The sets of winning states for Players~$\pz$~and~$\po$
are effectively computable as regular languages.
Furthermore, from each state, memoryless strategies suffice for the winning player.
\end{theorem}
We give the proof in several steps.
First, we show that the game induced by an SG-LCS contains a finite
attractor (Lemma~\ref{fattractors:lemma}).
Then, we show that the scheme in
Section~\ref{reachability:section} for computing winning states
wrt.\ reachability objectives
is guaranteed to terminate (Lemma~\ref{reachable:termination:lemma}).
Furthermore, we show that the scheme in
Section~\ref{parity:section} for computing winning states
wrt. \as{} parity objectives is guaranteed to terminate
(Lemma~\ref{parity:termination:lemma}).
Notice that 
Lemmas~\ref{reachable:termination:lemma}~and~\ref{parity:termination:lemma}
imply that for SG-LCS our transfinite constructions 
stabilize below $\omega$ (the first infinite ordinal).
Finally, we show that each step in the above two schemes 
can be performed using standard operations on regular languages
(Lemmas~\ref{reachability:compuability:lemma}~and~\ref{parity:compuability:lemma}).

\parg{Finite attractor.}
In \cite{Parosh:etal:attractor:IC} it was shown that
any Markov chain induced
by a Probabilistic LCS contains a finite attractor.
The proof can be carried over in a straightforward manner
to the current setting.
More precisely, the finite attractor is given by
$\attractor=(\lcsstates\times\emptychannels\times\{0,1\})$
where $\emptychannels(\channel)=\emptyword$ for each $\channel\in\channels$.
In other words, $\attractor$ is given by
the set of states in which all channels are empty.
  The proof relies on the observation that if the number of messages
  in some channel is sufficiently large, it is more likely that the number of
  messages decreases than that it increases in the next step.
This gives the following.
\begin{lemma}
\label{fattractors:lemma}
$\game$ contains a finite attractor.
\end{lemma}

\parg{Termination of Reachability Scheme.}

For a set of states $\stateset\subseteq\states$,
we  define the {\it upward closure} of $\stateset$ by
$\ucof\stateset:=\setcomp{\state}{\exists\state'\in\stateset.\,\state'\preceq\state}$.
A set $\ucset\subseteq\stateset\subseteq\states$
is said to be {\it $\stateset$-upward-closed}
(or {\it $\stateset$-u.c.\ } for short) if
$(\ucof{\ucset})\cap\stateset=\ucset$.
We say that $\ucset$ is {\it upward closed} if it is
$\states$-u.c.

\begin{lemma}
\label{higman:lemma}
  If $\stateset_0\subseteq \stateset_1\subseteq\cdots$, 
and for all $i$ it holds that
  $\stateset_i\subseteq \stateset$ and $\stateset_i$ is $\stateset$-u.c., then there is an $\tp\in\nat$
  such that $\stateset_i=\stateset_\tp$ for all $i\geq \tp$.
\end{lemma}

Now, we can show termination of the reachability scheme.
\begin{lemma}
\label{reachable:termination:lemma}
  There exists an $\tp\in\nat$ such
  that $\reachset_i=\reachset_\tp$ for all $i\geq\tp$.
\end{lemma}
\begin{proof}
First, we show that $\rcut{\reachset_i-\targetset}$ is 
$(\gcomplementof\game\targetset)$-u.c. for all $i\in\nat$.
We use induction on $i$.
For $i=0$ the result is trivial since
$\reachset_i-\targetset=\emptyset$.
For $i>0$, suppose that
$\state=(\lcsstate,\chassignment,0)\in\rcut{\reachset_i}-\targetset$.
This means that 
$\state\transition(\lcsstate,\chassignment',1)\in\reachset_{i-1}$
for some $\chassignment'\preceq\chassignment$, and hence
$\state'\transition(\lcsstate,\chassignment',1)$
for all $\state\preceq\state'$.

By Lemma~\ref{higman:lemma},
there is an $\tp'\in\nat$
  such that $\rcut{\reachset_i}-\targetset=\rcut{\reachset_{\tp'}}-\targetset$ 
 for all $i\geq \tp'$.
Since $\reachset_i\supseteq\targetset$ for all $i\geq 0$ it follows that 
 $\rcut{\reachset_i}=\rcut{\reachset_{\tp'}}$
 for all $i\geq \tp'$.
%

Since the graph of $\game$ is bipartite
(as explained in Section~\ref{sglcs:section}),
we have
$\xcut{\preof\game{\reachset_i}}=\xcut{\preof\game{\rcut{\reachset_i}}}$
and
$\ycut{\dualpreof\game{\reachset_i}}=\ycut{\dualpreof\game{\rcut{\reachset_i}}}$.
Since $\rcut{\reachset_i}=\rcut{\reachset_{\tp'}}$ for all $i\geq\tp'$,
we thus have
$\xcut{\preof\game{\reachset_i}}=\xcut{\preof\game{\rcut\reachset_{\tp'}}}\subseteq\reachset_{\tp'+1}$ and
$\ycut{\dualpreof\game{\reachset_i}}=\ycut{\dualpreof\game{\rcut\reachset_{\tp'}}}\subseteq\reachset_{\tp'+1}$.
It then follows that $\reachset_i=\reachset_{\tp}$ for all $i\geq\tp:=\tp'+1$.
\ignore{
Since the graph of $\game$ is bipartite (as explained in 
Section~\ref{sglcs:section}) it follows that either
$\rcut{\reachset_{i+1}}-\rcut{\reachset_i}=\reachset_{i+1}-\reachset_i$ or 
$\rcut{\reachset_{i+1}}-\rcut{\reachset_i}=\emptyset$
for all $i\geq 0$.
The result follows immediately.
}%
\end{proof}

\parg{Termination of Parity Scheme.}
We use several auxiliary lemmas.
The following lemma states that sink-freeness is preserved by the 
reachability scheme.
\begin{lemma}
\label{sink:free:sink:free:lemma}
If $\targetset$ is sink-free then $\xforceset(\game,\targetset)$
is sink-free.
\end{lemma}

\begin{lemma}
\label{sink:free:uc:lemma}
If  $\targetset$ is sink-free then 
$\rcut{\xforceset(\game,\targetset)}$ is upward closed.
\end{lemma}

\begin{lemma}
\label{sglcs:termination:lemma}
Let $\set{\stateset_i}_{i\in\ord}$ and 
$\set{\stateset'_i}_{i\in\ord}$ be sequences of sets of states
such that
(i)
Each $\stateset'_i$ is sink-free;
(ii)
$\stateset_i=\stateset'_i\cup\xforceset(\game,\bigcup_{j<i}\stateset_j)$;
(iii)
$\stateset'_i$ and $\xforceset(\game,\bigcup_{j<i}\stateset_j)$ 
are disjoint for all $i$.
Then, there is an $\tp\in\nat$ such that
$\stateset_i=\stateset_\tp$ for all $i\geq\tp$.
\end{lemma}

To apply Lemma~\ref{sglcs:termination:lemma}, we prove the following
two lemmas.
\begin{lemma}
\label{cset:trap:lemma}
$\cset_n(\game)$ is a $(\py)$-trap.
\end{lemma}
\begin{proof}
$\cset_0(\game)$ is trivially a $(\py)$-trap.
For  $i\geq 1$, the result follows immediately from
Lemma~\ref{tp:lemma} and 
Lemma~\ref{not:reach:trap:lemma}.
\end{proof}
\begin{lemma}
\label{cset:dset:sink:free:lemma}
For any game of rank $n$ both 
$\cset_n(\game)$ and $\dset_n(\game)$ are sink-free.
\end{lemma}
\begin{proof}

If $n=0$, then by definition $\cset_n(\game)=\dset_n(\game)=\states$,
which is sink-free by assumption.
Next, assume $n\geq 1$.
By Lemma~\ref{cset:trap:lemma} we know that $\cset_n(\game)$ is a $(\py)$-trap
and hence also sink-free.
  To prove the claim for $\dset_n(\game)$, we use induction on $i$ and prove
  that both $\uset_i$ and $\vset_i$ are sink-free.
Assume $\uset_j$ and $\vset_j$ are sink-free for all $j<i$.
  Then $\bigcup_{j<i}\vset_j$ is sink-free, and hence $\uset_i$
  is sink-free by Lemma~\ref{sink:free:sink:free:lemma}.
  Since $\cset_n(\game\cut\uset_i)$ and $\uset_i$ are sink-free, 
  it follows that $\vset_i$ is sink-free.
\end{proof}
Now, we apply Lemma~\ref{sglcs:termination:lemma}
 to prove that the sequences
$\set{\xset_i}_{i\in\ord}$ and
$\set{\uset_i}_{i\in\ord}$ terminate.
First, by Lemma~\ref{cset:dset:sink:free:lemma} we know that
$\dset_{n-1}(\game\cut\xset_i\cut\zset_i)$ is sink-free.
We know that
$\xset_i$ and $\dset_{n-1}(\game\cut\xset_i\cut\zset_i)$
are disjoint
since 
$\dset_{n-1}(\game\cut\xset_i\cut\zset_i)\subseteq
\gcomplementof\game{(\xset_i\cup\zset_i)}$.
Hence, we can  apply Lemma~\ref{sglcs:termination:lemma}
with $\stateset_i=\yset_i$,
$\stateset'_i=\dset_{n-1}(\game\cut\xset_i\cut\zset_i)$,
and conclude that $\set{\yset_i}_{i\in\ord}$ terminates, and hence
$\set{\xset_i}_{i\in\ord}$ terminates.
Second, by Lemma~\ref{cset:dset:sink:free:lemma} we know that
$\cset_{n}(\game\cut\uset_i)$ is sink-free.
Since 
$\cset_{n}(\game\cut\uset_i)\subseteq
\gcomplementof\game{\uset_i}$, we know that
$\uset_i$ and $\cset_{n}(\game\cut\uset_i)$
are disjoint.
Hence, we can  apply Lemma~\ref{sglcs:termination:lemma}
with $\stateset_i=\vset_i$,
$\stateset'_i=\cset_{n-1}(\game\cut\uset_i)$,
and conclude that $\set{\vset_i}_{i\in\ord}$ terminates, and hence
$\set{\uset_i}_{i\in\ord}$ terminates.
This gives the following lemma.
\begin{lemma}
\label{parity:termination:lemma}
There is an  $\tp\in\nat$ such that $\xset_i=\xset_\tp$ for all
$i\geq\tp$.
There is a  $\btp\in\nat$ such that $\uset_i=\uset_\btp$ for all
$i\geq\btp$.
\end{lemma}

\parg{Computability.}
For a given regular set $\reachset$,
the set $\preof\game\reachset$ 
is effectively regular \cite{Parosh:etal:CSL}, i.e.,
computable as a regular language.
The following lemma then follows from the fact that the other operations
used in computing $\xforceset(\game,\targetset)$
are those of set complement and union, which are effective 
for regular languages.
\begin{lemma}
\label{reachability:compuability:lemma}
If $\targetset$ is regular then
$\xforceset(\game,\targetset)$ is effectively regular.
\end{lemma}
\begin{lemma}
\label{parity:compuability:lemma}
For each $n$, both $\cset_n(\game)$ and 
$\dset_n(\game)$ are 
effectively regular.
\end{lemma}
\begin{proof}
The set $\states$ is regular, and hence 
$\cset_0(\game)=\dset_0(\game)=\states$
is effectively regular.
The result for $n>0$ follows from
Lemma~\ref{reachability:compuability:lemma} and from the fact
that the rest of the operations used to build  $\cset_n(\game)$ and 
$\dset_n(\game)$ are those of set complement and union.
\end{proof}
\parg{Remark.}
Although we use Higman's lemma for showing termination  of 
our fixpoint computations, our proof differs significantly
from the standard ones for well quasi-ordered transition systems
\cite{Parosh:Bengt:Karlis:Tsay:general:IC}.
For instance, the generated sets are in general
not upward closed wrt. the underlying
ordering $\preceq$.
Therefore, we need to use the notion of $\stateset$-upward closedness for a set
of states $\stateset$.
More importantly, we need to define new 
(and much more involved) sufficient conditions
for the termination of the computations
(Lemma~\ref{sglcs:termination:lemma}), and to show that these conditions
are satisfied
(Lemma~\ref{cset:dset:sink:free:lemma}).

\section{Conclusions and Discussion}
\label{conclusions:section}
\begin{figure}
\begin{tikzpicture}[show background rectangle]
\node[name=dummy] {};
\node[prob-node,name=s0] at (dummy) {$s_0$};
\node[prob-node,name=s1,anchor=west] at ($(s0.east)+(3mm,0mm)$) {$s_1$};
\node[prob-node,name=s2,anchor=west] at ($(s1.east)+(3mm,0mm)$) {$s_2$};
\node[prob-node,name=s3,anchor=west] at ($(s2.east)+(3mm,0mm)$) {$s_3$};
\node[name=s4,anchor=west] at ($(s3.east)+(2mm,0mm)$) {$\cdots$};

\node[anchor=south] at ($(s4.north)+(-5mm,5mm)$) {(a)};

\draw[transition-edge] (s0) to [out=150,in=210,loop] node[above,near start]{\scriptsize $0.3$} (s0);
 
\draw[transition-edge] (s0) to [out=60,in=120] node[above]{\scriptsize $0.7$} (s1);
\draw[transition-edge] (s1) to [out=240,in=300] node[below]{\scriptsize $0.3$} (s0);

\draw[transition-edge] (s1) to [out=60,in=120] node[above]{\scriptsize $0.7$} (s2);
\draw[transition-edge] (s2) to [out=240,in=300] node[below]{\scriptsize $0.3$} (s1);

\draw[transition-edge] (s2) to [out=60,in=120] node[above]{\scriptsize $0.7$} (s3);
\draw[transition-edge] (s3) to [out=240,in=300] node[below]{\scriptsize $0.3$} (s2);

\end{tikzpicture}
{ }
\begin{tikzpicture}[show background rectangle]
\node[name=dummy0] {};
\node[player-node,name=s0] at (dummy0) {$s_0$};
\node[name=dummy1,anchor=north] at ($(dummy0.south)+(0mm,-13mm)$) {};
\node[prob-node,name=s2,anchor=east] at ($(dummy1.west)+(-3mm,0mm)$) {$s_2$};
\node[prob-node,name=s1,anchor=east] at ($(s2.west)+(-6mm,0mm)$) {$s_1$};
\node[prob-node,name=s3,anchor=west] at ($(dummy1.east)+(3mm,0mm)$) {$s_3$};
\node[prob-node,name=s4,anchor=west] at ($(s3.east)+(6mm,0mm)$) {$s_4$};
\node[name=s5,anchor=west] at ($(s4.east)+(2mm,0mm)$) {$\cdots$};

\node[anchor=south] at ($(s5.north)+(-5mm,10mm)$) {(b)};

\draw[transition-edge] (s1) to [out=150,in=210,loop] node[above,near start]{$1$} (s1);
 
\draw[transition-edge] (s0) to [out=150,in=80] (s1);

\draw[transition-edge] (s0) to [out=240,in=100] (s2);
\draw[transition-edge] (s2) to [out=50,in=260] node[below=5pt]{\scriptsize $0.5$} (s0);

\draw[transition-edge] (s0) to [out=310,in=70] (s3);
\draw[transition-edge] (s3) to [out=130,in=280] node[below=5pt]{\scriptsize $0.5$} (s0);

\draw[transition-edge] (s0) to [out=30,in=100] (s4);
\draw[transition-edge] (s4) to [out=160,in=340] node[above=5pt]{\scriptsize $0.5$} (s0);

\draw[transition-edge] (s4) to [] node[below]{\scriptsize $0.5$} (s3);
\draw[transition-edge] (s3) to [] node[below]{\scriptsize $0.5$} (s2);
\draw[transition-edge] (s2) to [] node[below]{\scriptsize $0.5$} (s1);

\end{tikzpicture}

\caption{ (a) Finite attractor requirement.
(b) Finite strategy requirement.}
\label{req:figure}
\end{figure}
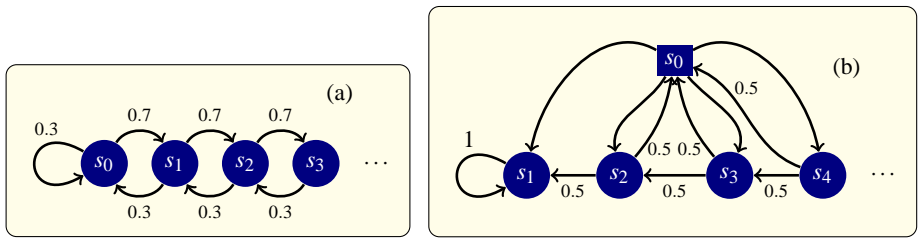

We have presented a scheme for solving stochastic games with \as{} parity winning
conditions  under the two requirements that
(i)  the game contains a finite
attractor and 
(ii) both players are restricted to finite-memory strategies.
We have shown that this class of games is memoryless determined.
The method is instantiated to prove decidability
of \as{} parity games induced by lossy channel systems.
The two above requirements  are both necessary for
our method.
To see why our scheme fails if the game lacks a {\bf finite attractor},
consider the game in Figure~\ref{req:figure}~(a)
 (a variant of the Gambler's ruin problem).
All states are random, i.e., $\zstates=\ostates=\emptyset$, 
and $\coloring(\state_0)=1$ and $\coloring(\state_i)=0$ when $i>0$. 
The probability to go right from
any state is $0.7$ and the probability to go left (or to make a
self-loop in $\state_0$) is $0.3$.
This game does not have any finite attractor. 
It can be shown that
the probability to reach $\state_0$ infinitely often is 0 for all initial
states. However, our construction will classify all states as winning
for player 1.
More precisely, the construction of $\cset_1(\game)$ converges
after one iteration with $\zset_i=S, \xset_i=\emptyset$ for all $i$ and
$\cset_1(\game)=S$.
Intuitively, the problem is that even if the force-set of $\set{\state_0}$ 
(which is the entire set of states) is visited infinitely many times, 
the probability of visiting $\set{\state_0}$ infinitely often is still zero,
since the probability of returning to $\set{\state_0}$ gets smaller and
smaller. Such behavior is impossible in a game graph that contains
a finite attractor.

We restrict both players to {\bf finite-memory strategies}. This is a
different problem from when arbitrary strategies are allowed (not a
sub-problem). 
In fact, it was shown in \cite{BBS:ACM2007} that
for arbitrary strategies, the problem is undecidable.
Figure~\ref{req:figure}~(b)
gives an example of a game graph where the two problems yield
different results (see also \cite{BBS:ACM2007}).
Player 1 controls $\state_0$, whereas $\state_1, \state_2, \dots$ are
random; $\coloring(\state_0)=0, \coloring(\state_1)=2,
\coloring(\state_i)=1$ if $i\geq 2$.
The transition probabilities are $\prob(\state_1,\state_1)=1$ and
$\prob(\state_n,\state_{n-1})=\prob(\state_n,\state_0)=\frac12$ when $n\geq
2$.
Player 1 wants to ensure that the highest color that is seen infinitely
often is odd, and thus wants to avoid state $s_1$ (which has color 2).
If the players can use arbitrary strategies, then although player 1 cannot win
with probability 1, he can win with a probability arbitrarily close
to 1 using an infinite-memory strategy: player 1 goes from
$\state_0$ to $\state_{k+i}$ when the play visits $\state_0$ for the
$i$'th time. Then player 1 wins with probability 
$\prod_{i=1}^{\infty} (1-2^{-k-i+1})$, which can be made arbitrarily close to
$1$ for sufficiently large $k$.
In particular, player 0 does not win \as{} in this case.
On the other hand, if the players are limited to finite-memory
strategies, then no matter what strategy player 1 uses, the play
visits $\state_1$ infinitely often with probability 1, so player 0
wins almost surely; this is also what our algorithm computes.

As future work, we will consider extending our framework
to (fragments of) probabilistic extensions of other models
such as Petri nets and noisy Turing machines
\cite{Parosh:etal:MC:infinite:journal}.

\hide{There exist more general winning conditions; in particular, the Muller
condition supersedes the parity, Rabin, and Streett conditions.
There is a known reduction from Muller conditions to parity conditions,
where the game graph is transformed using a data structure introduced
in \cite{GuHa:stoc92}.
We believe that the combination of our method with this reduction would
make it possible to determine the winner in stochastic Muller games played
on lossy channel systems, but this remains as future work.
}

\def\Nst#1{$#1^{st}$}\def\Nnd#1{$#1^{nd}$}\def\Nrd#1{$#1^{rd}$}\def\Nth#1{$#1^{th}$}


\newpage
\appendix
\section{Proofs of Lemmas}
\label{proofs:section}

\subsection*{Lemma~\ref{trap:certainly:lemma}}

We define a memoryless strategy $\xstrat$ of Player~$\px$ that is 
surely winning
from any state $\state\in\stateset$, i.e.,
$\stateset\subseteq\xvinset(\xstrat,\yallstrats(\game))(\game,\always\stateset)$.
For a state $\state\in\xcut\stateset$, we define
$\xstrat(\state)=\selectfrom{\postof\game{\state}\cap\stateset}$.
This is well-defined since $\stateset$
is a $(\py)$-trap.
We can now show that any run that starts
from a state $\state\in\stateset$ and
that is consistent with 
$\xstrat$ will surely remain inside
$\stateset$.
Let $\ystrat$ be any strategy of Player~$\py$, and
let $\state_0,\state_1,\ldots\in\runsof{\game,\state,\xstrat,\ystrat}$.
We show, by induction on $i$,
that $\state_i\in\stateset$ for all $i\geq 0$.
The base case is clear since $\state_0=\state\in\stateset$.
For $i>1$, we consider three cases depending on $\state_i$:
\begin{itemize}
\item
$\state_i\in\xcut\states$.
By the induction hypothesis we know that 
$\state_i\in\stateset$,
and hence by definition of $\xstrat$ we know that 
$\state_{i+1}=\xstrat(\state_i)\in\stateset$.
\item
$\state_i\in\ycut\states$.
By the induction hypothesis we know that 
$\state_i\in\stateset$,
and hence $\state_{i+1}\in\stateset$
since $\stateset$ is a $(\py)$-trap.
\item
$\state_i\in\rcut\states$.
By the induction hypothesis we know that 
$\state_i\in\stateset$,
and hence $\state_{i+1}\in\stateset$
since $\stateset$ is closable. 

\end{itemize}

\subsection*{Lemma~\ref{not:reach:trap:lemma}}

By Lemma~\ref{reachability:tp:lemma} and the definition of
$\yavoidset(\game,\targetset)$ it follows that \\
$\yavoidset(\game,\targetset)=\gcomplementof\game{\reachset_\tp}$ and
that $\reachset_{\tp+1}\subseteq\reachset_\tp$.
First, we show that $\gcomplementof\game\reachset_\tp$ 
is sink-free as follows.
For a state $\state\in\states$, we show that
$\postof\game\state\cap
(\gcomplementof\game{\reachset_\tp})\neq\emptyset$.
There are three cases to consider.
\begin{itemize}
\item
$\state\in\xcut{\gcomplementof\game\reachset_\tp}$.
Suppose that $\postof\game\state\not\subseteq(\gcomplementof\game\reachset_\tp)$.
It follows that $\postof\game\state\cap\reachset_\tp\neq\emptyset$.
Hence, $\state\in\reachset_{\tp+1}\subseteq\reachset_\tp$
which is a contradiction.
This means that
$\postof\game\state\subseteq\gcomplementof\game\reachset_\tp$.
Since $\states$ is sink-free, we know that
$\postof\game\state\neq\emptyset$.
Consequently, $\postof\game\state\cap\gcomplementof\game\reachset_\tp\neq\emptyset$.
\item
$\state\in\ycut{\gcomplementof\game\reachset_\tp}$.
Suppose that $\postof\game\state\subseteq\reachset_\tp$.
It follows that $\state\in\reachset_{\tp+1}\subseteq\reachset_\tp$
which is a contradiction.
Hence,
$\postof\game\state\cap\gcomplementof\game\reachset_\tp\neq\emptyset$.
\item
$\state\in\rcut{\gcomplementof\game\reachset_\tp}$.
The claim follows in similar manner to the case where 
$\state\in\xcut{\reachset_\tp}$.
\end{itemize}
Second, when proving sink-freeness above,  we showed that
$\postof\game\state\subseteq\gcomplementof\game\reachset_\tp$
for any $\state\in\rcut{\gcomplementof\game\reachset_\tp}$ 
which means that $\gcomplementof\game\reachset_\tp$  
is closable.
Finally, when proving sink-freeness, we also showed
that $\postof\game\state\subseteq\gcomplementof\game\reachset_\tp$
for any $\state\in\xcut{\gcomplementof\game\reachset_\tp}$
which completes the proof.

\subsection*{Lemma~\ref{reachability:correct:lemma}}

Let $\reachset=\xforceset(\game,\targetset)$.
To prove the first claim, we define a memoryless  strategy $\xstrat$ of Player~$\px$ that is winning
from any state $\state\in\reachset$, i.e.,
$\reachset\subseteq\xwinset(\xstrat,\yallstrats)(\game,\eventually\targetset^{>0})$.
For any $\state\in\xcut{\reachset_{i+1}-\reachset_i}$ 
where $i+1$ is a successor
ordinal, we define 
$\xstrat(\state):=\selectfrom{\postof\game{\state}\cap\reachset_i}$.
We show that $\xstrat$ is a winning strategy
for Player~$\px$.
Fix a strategy $\ystrat$ for Player~$\py$.
We show that 
$\probm_{\game,\state,\xstrat,\ystrat}(\eventually\targetset)>0$.
We prove the claim using transfinite induction.
If $\state\in\reachset_0$ then the claim follows trivially.
If $\state\in\reachset_{i+1}$ where $i+1$ is
a successor ordinal then either
$\state\in\reachset_{i}$
in which case the claim holds by the induction hypothesis,
or $\state\in\reachset_{i+1}-\reachset_{i}$.
We consider three cases:
\begin{itemize}
\item
  $\state\in\xcut{\reachset_{i+1}-\reachset_i}$.
  By definition of $\xstrat$, 
  we know that $\xstrat(\state)=\state'$ for some $\state'\in\reachset_{i}$.
  By the induction hypothesis we know that 
  $\probm_{\game,\state',\zstrat,\ostrat}(\eventually\targetset)>0$ and hence
  $\probm_{\game,\state,\zstrat,\ostrat}(\eventually\targetset)>0$.
\item
  $\state\in\ycut{\reachset_{i+1}-\reachset_{i}}$.
  Let $\ystrat(\state)=\state'$.
  By definition of $\reachset_{i+1}$
  we know that $\state'\in\reachset_{i}$.
  Then, the proof follows as in the previous case.
\item
  $\state\in\rcut{\reachset_{i+1}-\reachset_{i}}$.
  By definition of $\reachset_{i+1}$ we know that 
  there is a $\state'\in\reachset_{i}$
  such that $\prob(\state,\state')>0$.
  By the induction hypothesis, it follows that 
  $\probm_{\game,\state,\zstrat,\ostrat}(\eventually\targetset)
  \geq 
  \probm_{\game,\state',\zstrat,\ostrat}(\eventually\targetset)\cdot\prob(\state,\state')>0$.
\end{itemize}
Finally, if $\state\in\reachset_{i}$ where $i>0$ is
a limit ordinal, then we know that
$\state\in\reachset_{j}$ 
for some $j<i$.
The claim then follows by the induction hypothesis.

From Lemma~\ref{not:reach:trap:lemma} and
Lemma~\ref{trap:certainly:lemma} it follows that
there is a strategy $\ystrat$ for Player~$\py$ such that
$\xavoidset(\game,\targetset)\subseteq\yvinset(\ystrat,\xallstrats)(\game,\always(\xavoidset(\game,\targetset)
))$.
The second claim follows then from the fact that 
$\targetset\cap \xavoidset(\game,\targetset)=\emptyset$.

\subsection*{Lemma~\ref{higman:lemma}}

  By Higman's lemma \cite{higman:divisibility}, there is a $\tp\in\nat$ such that
  $\stateset_i\uparrow=\stateset_\tp\uparrow$ for all $i\geq \tp$.
  Hence, $\stateset_i\uparrow\cap \stateset=\stateset_\tp\uparrow\cap \stateset$
  for all $i\geq\tp$.
  Since all $\stateset_i$ are $\stateset$-u.c.,
  $\stateset_i\uparrow\cap \stateset=\stateset_i$
  for all $i\geq\tp$.
  So $\stateset_i=\stateset_\tp$ for all $i\geq \tp$.

\subsection*{Lemma~\ref{sink:free:sink:free:lemma}}

  We prove by induction on $i$ that $\reachset_i$ is sink-free for all
  ordinals $i\in\ord$.
  If $i=0$, the claim holds by assumption.
  If $i+1$ is a successor ordinal, then the claim follows from the
  definition of $\reachset_{i+1}$ (all new states in $\reachset_{i+1}$ have a
  successor in $\reachset_i$).
  If $i>0$ is a limit ordinal, then each state in $\reachset_i$ belongs to
  some $\reachset_j$ with $j<i$ for which the claim holds by the induction
  hypothesis.

\subsection*{Lemma~\ref{sink:free:uc:lemma}}

  Take any $\state\in\rcut{\xforceset(\game,\targetset)}$ 
and take $\state'\in \states$ such that
  $\state\preceq \state'$.
  By Lemma~\ref{sink:free:sink:free:lemma}, $\state$ has a successor $\state''\in\xforceset(\game,\targetset) $.
  Thus, $\state''\in \reachset_i$ for some $i\in\ord$.
  Since $\game$ is induced by a SG-LCS, $\state'$ is a predecessor of every
  successor of $\state$ (including $\state''$).
  Hence, $\state'\in \reachset_{i+1}$.

\subsection*{Lemma~\ref{sglcs:termination:lemma}}

First, define $F_i:=\xforceset(\game,\bigcup_{j<i}\stateset_j)$.
We perform the proof in several steps:
(i) $F_0\subseteq\stateset_0\subseteq F_1\subseteq \stateset_1\subseteq\cdots$.
This follows from the fact that 
$\stateset_i=\stateset'_i\cup F_i$
and the definition of $F_i$.
(ii) There is an
$\tp\in\nat$ such that $\rcut{F_i}=\rcut{F_\tp}$ 
for all $i\geq\tp$.
By Lemma~\ref{sink:free:sink:free:lemma}
(and induction on $i$),
each $F_i$ is sink-free and hence
by Lemma~\ref{sink:free:uc:lemma} 
each $\rcut{F_i}$ is upward closed.
By Lemma~\ref{higman:lemma} it follows that there is a
$\tp\in\nat$ such that $\rcut{F_i}=\rcut{F_\tp}$ 
for all $i\geq\tp$.
%
(iii)
$\rcut{\stateset_i}=\rcut{F_i}=\rcut{\stateset_\tp}$ for all $i\geq\tp$.
  This follows from the previous two steps.
%
(iv) $\rcut{\stateset'_i}=\emptyset$ for all $i\geq \tp+1$.
From the assumption that
$\stateset_i$ is the disjoint union of $\stateset'_i$ and $F_i$,
$\stateset'_i=\stateset_i-F_i$.
By the previous step, this set is empty.
(v) $\stateset'_i=\emptyset$ for all $i\geq \tp+1$.
    Since the transition system generated by an SG-LCS is bipartite
    with every second state in $\rstates$, and since every member of a sink-free
    set has a successor in the set itself, it follows that any nonempty sink-free set
    contains both states in $\rstates$ and states that belong to one of 
the players~$\pz$~or~$\po$.
    Since $\stateset'_i$ is sink-free (by assumption) but does not contain any
    states in $\rstates$ (by the result in previous step), $\stateset'_i$
    has to be
    empty.
(vi)
$\stateset_i=\stateset_{\tp}$ for all $i\geq \tp$.
    By the previous step, we know that 
$\stateset_i=\xforceset(\game,\bigcup_{j<i}\stateset_j)$ for all
    $i\geq \tp+1$.
By the definition of $\xforceset$ it follows that $\stateset_i=\stateset_{\tp}$ for all $i\geq \tp+1$.

\end{document}